\documentclass[aip,amsmath,amssymb,reprint]{revtex4-1}

\usepackage{graphicx}
\usepackage{dcolumn}% Align table columns on decimal point
\usepackage{bm}% bold math
\usepackage{upgreek}
\usepackage[utf8]{inputenc}
\usepackage[T1]{fontenc}
\usepackage{mathptmx}
\usepackage{etoolbox}
\usepackage{color}

\makeatletter
\def\@email#1#2{%
 \endgroup
 \patchcmd{\titleblock@produce}
  {\frontmatter@RRAPformat}
  {\frontmatter@RRAPformat{\produce@RRAP{*#1\href{mailto:#2}{#2}}}\frontmatter@RRAPformat}
  {}{}
}%
\makeatother

\begin{document}

%\preprint{AIP/123-QED}

% [Molecular Auger Decay with Complex-Variable Methods]
\title{Molecular Auger Decay Rates from Complex-Variable Coupled-Cluster Theory}
\author{Florian Matz}
\email{thomas.jagau@kuleuven.be}
\author{Thomas-C. Jagau}

\email{florian.matz@kuleuven.be}
\affiliation{Division of Quantum Chemistry and Physical Chemistry, 
KU Leuven, Celestijnenlaan 200F, 3001 Leuven, Belgium}

\date{\today} 

\begin{abstract} \setlength{\parindent}{0pt}
The emission of an Auger electron is the predominant relaxation mechanism of 
core-vacant states in molecules composed of light nuclei. In this non-radiative 
decay process, one valence electron fills the core vacancy while a second valence 
electron is emitted into the ionization continuum. Because of this coupling to 
the continuum, core-vacant states represent electronic resonances that can be 
tackled with standard quantum-chemical methods only if they are approximated 
as bound states, meaning that Auger decay is neglected. 

Here, we present an approach to compute Auger decay rates of core-vacant states 
from coupled-cluster and equation-of-motion coupled-cluster wave functions combined 
with complex scaling of the Hamiltonian or, alternatively, complex-scaled basis 
functions. Through energy decomposition analysis, we illustrate 
how complex-scaled methods are capable of describing the coupling to the ionization 
continuum without the need to model the wave function of the Auger electron explicitly. 
In addition, we introduce in this work several approaches for the determination of 
partial decay widths and Auger branching ratios from complex-scaled coupled-cluster 
wave functions. 

We demonstrate the capabilities of our new approach by computations on core-ionized 
states of neon, water, dinitrogen, and benzene. Coupled-cluster and equation-of-motion 
coupled-cluster theory in the singles and doubles approximation both deliver excellent 
results for total decay widths, whereas we find partial widths more straightforward 
to evaluate with the former method. We also observe that the requirements towards 
the basis set are less arduous for Auger decay than for other types of resonances 
so that extensions to larger molecules are readily possible. 

% In this work, we develop an efficient and precise computational description of these states that overcomes the challenges they pose concerning electron correlation and coupling to the continuum. The method of complex scaled basis functions, together with the equation of motion-coupled-cluster approach, results in very accurate values for partial widths when we allow any complex number for the scaling factor. This corresponds with an optimization of the exponents of the scaled functions, which always results in virtual orbitals that overlap with the valence orbitals.  Thus, the only requirement on basis sets used for these calculations is the accurate description of the valence states and useful results can be obtained with basis sets smaller than aug-cc-pCVTZ. Employing the basis functions optimized following our method is successful for the description of the resonant core-ionized state of neon atoms as well as water, nitrogen, and benzene molecules. The resulting wave functions can be used to obtain the contributions of specific decay channels to the decay rate by projecting out valence-core excitations from the density matrices. 

\end{abstract}

\maketitle

%%%%%%%%%%%%%%%%%%%%%%%%%%%%%%%%%%%%%%%%%%%%%%%%%%%%%%%%%%%%%%%%%%%%%%%%%%%%%%%%%%%

\section{Introduction}
X-ray spectroscopy is a valuable tool for the analysis of structure and reactivity 
throughout chemistry.\cite{agarwal13} Not only does the complexity and accuracy 
of experiments advance every year, but this has also entailed growing interest 
in theoretical modeling of the interaction of atoms and molecules with X-ray 
radiation and the resulting core-vacant states.\cite{norman18,zimmermann20} 
Experiment and theory strongly rely on each other for the examination of 
systems with core vacancy; in many cases, the explanation and interpretation 
of experimental results requires input from theoretical modeling. At the same 
time, the unique electronic structure of core-vacant states poses a challenge 
for theory. The variety of recent investigations illustrates the efforts 
to achieve a quantitatively correct and at the same time computationally 
affordable description of core-vacant states; overviews are available from 
Refs. \citenum{norman18,zimmermann20}.

\begin{figure*}
\centering
\includegraphics[width=.58\linewidth]{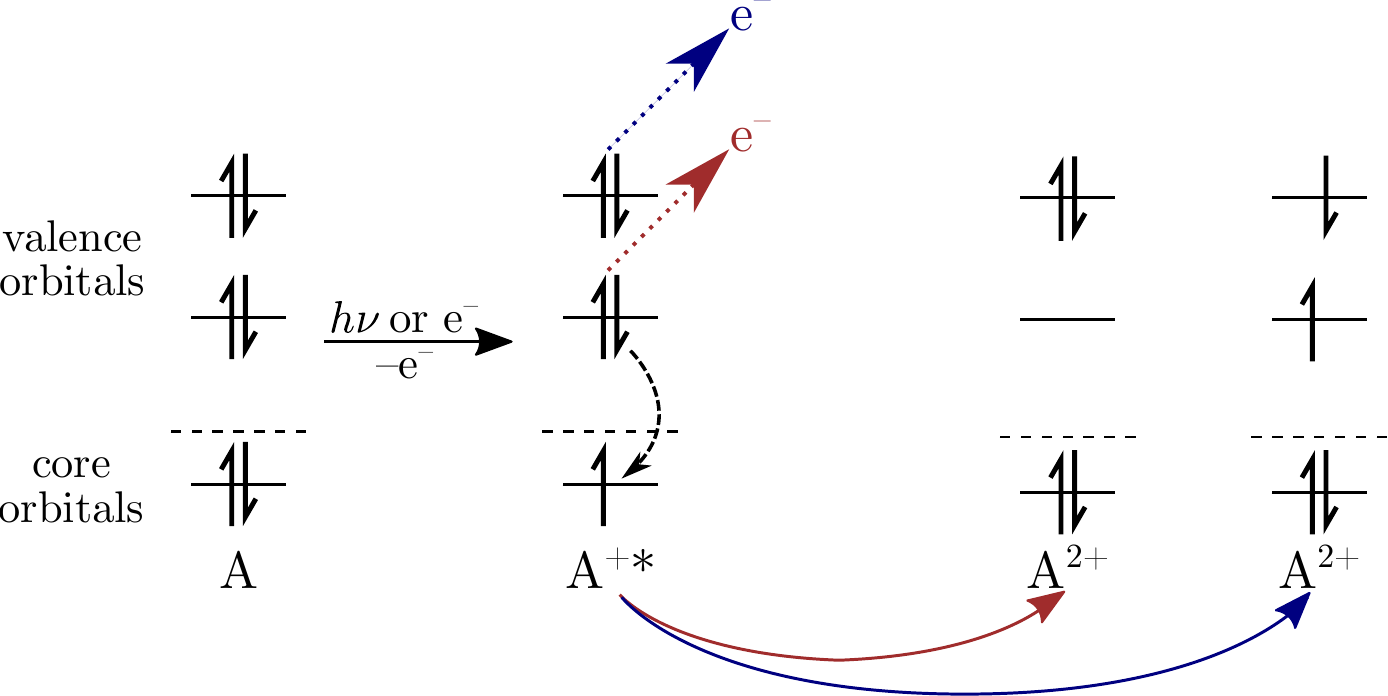}
\caption{Generation of a core-ionized state (left) and its Auger decay into 
different channels (right).}
\label{fig:auger}
\end{figure*}

X-ray irradiation of a neutral species can create both core-excited and core-ionized 
states. An important mechanism by which these highly excited states can relax is 
the Auger-Meitner effect,\cite{meitner22,auger23} a non-radiative decay process 
involving two valence electrons: One of them is emitted while the other one fills 
the core vacancy. Auger decay exists in several variants and can occur as a result 
of both core-ionization and core-excitation. As shown in Fig. \ref{fig:auger}, 
decay of a core-ionized state A$^{+*}$ produces the dication A$^{2+}$ in different 
electronic states, which are referred to as decay channels. The corresponding decay 
process of a neutral core-excited state is called resonant Auger decay.\cite{brown80,
armen00} Further variants are processes where cations with charges
higher than 2 are produced through simultaneous emission of multiple electrons, 
this has been demonstrated experimentally for double Auger decay~\cite{carlson65}
and triple Auger decay~\cite{muller15}. An exotic phenomenon is three-electron
Auger decay, in which two electrons simultaneously fill a double vacancy.~\cite{lee93,
feifel16} It is also common that the target states of Auger decay are subject 
to further decay resulting in so-called Auger cascades.\cite{agarwal13} Moreover, 
there are non-local decay processes such as interatomic and
intermolecular Coulombic decay\cite{cederbaum97,jahnke20} (ICD) and 
electron-transfer mediated decay. \cite{zobeley01} 
%which are of special interest
%to biochemistry since they act as a source of low-energy electrons in biological
%systems.\cite{stoychev11,hergenhahn12} 

The main subject of the present work is the description of Auger decay of 
core-ionized states, but many of our conclusions hold for resonant Auger decay 
and more involved processes as well. A particular topic that we will deal with 
is the determination of partial decay widths, that is, the relative probability 
of decay into a particular channel. In Auger electron spectroscopy,\cite{agarwal13} 
partial decay widths are determined from the kinetic energies of the emitted Auger 
electrons. The intensity of the Auger electrons is measured as a function of their 
energy and, typically, each signal can be assigned to a specific decay channel. 
While the total number of open decay channels can be anticipated by the application 
of selection rules based on molecular symmetry, the determination of partial decay 
widths poses a challenge to experiment and theory alike.~\cite{agarwal13,manne85,
zaehringer92a,zaehringer92b,tarantelli94,yarzhemsky02,kolorenc11,inhester12,
inhester14,skomorowski21a,skomorowski21b} In a rigorous scattering
approach, they may potentially be evaluated by constructing true continuum functions for
each channel at the respective energies. However, we consider it desirable to evaluate
partial widths from $L^2$ integrable wave functions in analogy to molecular properties
of bound states, even though this necessarily constitutes an approximation.

A fundamental aspect of core-vacant states is that they are not bound states 
but metastable electronic resonances.\cite{moiseyev11,jagau17} Since they 
can undergo Auger decay, these states are coupled to the continuum and their 
lifetime is finite. This is beyond the reach of quantum-chemical methods geared 
towards bound states. Many existing computational approaches for core-vacant 
states neglect their metastable character entirely, meaning the decay width 
is modeled to be zero. An elegant way to impose this restriction in a controlled 
manner consists in the core-valence separation (CVS).\cite{cederbaum80} There 
is ample evidence~\cite{coriani15,vidal19,fransson21} that CVS-based descriptions 
are highly accurate for many types of core-vacant states as long as one is only 
interested in energies and molecular properties determined as energy derivatives. 
However, methods that consider core-vacant states to be bound are obviously 
unsuited for modeling Auger decay. 

There are several theoretical approaches for electronic resonances.\cite{jagau17,
moiseyev11} Besides approaches based on R-matrix theory,\cite{garcia09,gorczyca00} 
Fano's theory \cite{fano61,feshbach62} is of particular importance for Auger 
decay. Here, the resonance wave function is modeled as a bound state superimposed 
by the electronic continuum. An important aspect of methods based on Fano's 
theory is that the electronic continuum cannot be properly represented by the 
$L^2$ integrable functions used in bound-state electronic-structure theory. 
This can be circumvented by modeling the wave function of the emitted Auger 
electron in an implicit fashion, for example, using Stieltjes 
imaging.\cite{langhoff74,carravetta87} Although the problem of a somewhat
arbitrary partition of the Hilbert space into a bound and a continuum part
persists, this approach is well capable of modeling Auger decay as the popular 
``Fano-ADC'' approach\cite{averbukh05,kolorenc20} illustrates. Here, algebraic
diagrammatic construction\cite{schirmer82} (ADC) is used as electronic-structure 
backend. As an alternative to Stieltjes imaging, more explicit treatments of 
the Auger electron are possible where one solves an effective one-electron 
Schrödinger equation with scattering boundary conditions.\cite{zaehringer92a,
zaehringer92b} Recently, an approach was suggested, where the decaying resonance 
state is represented as a product of a continuum orbital and a correlated 
many-body wave function.\cite{skomorowski21a} This approach has been shown 
to reproduce the most important signals in Auger spectra reliably, but the 
functional form of the wave function of the Auger electron has to be assumed 
\textit{a priori}, for example, as plane wave or Coulomb wave.

In this work, we seek to put forward an alternative method for the computation 
of molecular Auger decay widths based on an $L^2$ representation of the resonance 
wave function. This relies on analytic continuation of the Hamiltonian to the 
complex plane by means of complex scaling\cite{aguilar71,balslev71,moiseyev11} 
and its extension to molecules based on complex basis functions.\cite{mccurdy78,
moiseyev79} In these methods, the decaying character of the resonance states is
implicitly considered in $L^2$ integrable wave functions that are eigenstates 
of a non-Hermitian Hamiltonian and have complex energies. In this way, 
complex-variable electronic-structure methods offer a unified treatment 
of bound states and different types of resonances. Their integration into 
existing implementations of quantum-chemical methods\cite{jagau17,bravaya13,
jagau14,zuev14,white15a,white15b,white17} requires extension of the arithmetic 
to complex numbers and a different normalization of the wave function\cite{
moiseyev78} but the working equations of a particular quantum-chemical method 
stay the same and no \textit{a priori} assumption about the wave function of 
the emitted electron needs to be made. 

Complex scaling has already been applied to atomic Auger decay,\cite{zhang12a,
zhang12b,peng16} but no similar applications to molecules have been reported. 
At the same time, several applications of complex basis functions to other 
types of resonances such as temporary anions\cite{white15a,white15b,white17} 
and molecules in static electric fields have been reported recently.\cite{
jagau18,thompson19,hernandez19,hernandez20} We also mention treatments of 
Auger decay rates and ICD rates based on complex absorbing potentials.\cite{
ghosh13,ghosh15,ghosh17} This technique affords a treatment 
of electronic resonances in terms of $L^2$ wave functions as well and can be 
related to complex scaling.\cite{moiseyev98,riss93,riss98}

Here, we extend the method of complex basis functions to molecular Auger 
decay. Through energy decomposition analysis of the complex-scaled wave 
function, we identify key differences between core-ionized states and other 
types of resonances that involve only valence electrons. These differences 
give rise to markedly different basis requirements and an overall more robust 
performance of complex-scaled methods for core-ionized states as compared 
to other types of resonances. Our work is based on coupled-cluster\cite{
cizek66,cizek69,shavitt09} (CC) and equation-of-motion (EOM)-CC\cite{emrich81,
sekino84,stanton93a,nooijen93,stanton94,shavitt09} wave functions within the 
singles and doubles approximation (CCSD and EOM-CCSD). These methods provide 
a parameter-free single-reference description of the many-electron wave 
function. Several applications to X-ray spectroscopies\cite{coriani15,vidal19,
skomorowski21a,skomorowski21b,zheng19,park19,frati19,liu19,nanda20,vidal20a,vidal20b,
matthews20} have illustrated that they are able to provide an excellent 
description of core-vacant states. However, we anticipate that the analysis 
of complex-scaled CCSD and EOM-CCSD wave functions presented here will be 
relevant to other state-of-the-art electronic-structure methods for core-vacant 
states as well. These include, for example, time-dependent density functional 
theory (TD-DFT)~\cite{besley10} and $\Delta$DFT approaches,\cite{besley09} 
ADC methods of second and third order\cite{wenzel14a,wenzel14b,wenzel15,
averbukh05,kolorenc20}, and higher-order CC methods.\cite{zheng19,liu19,matthews20}

The remainder of this article is structured as follows: In Sec.~\ref{sec:theo}, 
we discuss the theory of complex scaling and complex basis functions, some aspects 
of complex-scaled CC and EOM-CC theory relevant to our work, and the theoretical 
background of our energy decomposition analysis. In Sec. \ref{sec:atres} we 
analyze the complex-scaled wave function of Ne$^+$ ($1\text{s}^{-1}$) and discuss the 
implications for the treatment of molecular Auger decay in terms of complex basis
functions. On the basis of these results, we present in Sec. \ref{sec:molres} a
computational protocol for the treatment of molecules together with 
some applications to core-ionized states of H$_2$O, N$_2$, and C$_6$H$_6$.
Our general conclusions and an outlook on possible extensions of the new method
are given in Sec. \ref{sec:conc}.

%%%%%%%%%%%%%%%%%%%%%%%%%%%%%%%%%%%%%%%%%%%%%%%%%%%%%%%%%%%%%%%%%%%%%%%%%%%%%%

\section{Theoretical considerations}
\label{sec:theo}

\subsection{Treatment of the continuum by means of complex scaling}
\label{sec:thcs}
In complex scaling (CS) \cite{aguilar71,balslev71,moiseyev11}, the Hamiltonian
is subject to an unbounded similarity transformation
\begin{equation} \label{eq:cs1}
\Hat{H}_\text{CS} = \Hat{S}\Hat{H}\Hat{S}^{-1}\text{ with }\Hat{S} = 
\text{e}^{\text{i}\theta r \, \text{d}/\text{d}r}~, \quad 
0 < \theta < \uppi/4 ~.
\end{equation}
This is equivalent to rotating the electronic coordinates in $\Hat{H}(r)$ 
so that the Hamiltonian becomes $\Hat{H}(r\text{e}^{\text{i}\theta})$. The 
resonances, which are peaks in the density of continuum states in Hermitian 
quantum mechanics, now attain discrete complex eigenvalues
\begin{equation} \label{eq:cs2}
E_\text{res} = E_\text{R}-\text{i}\Gamma/2,
\end{equation}
which are directly related to the resonance position $E_\text{R}$ and the 
resonance width $\Gamma$, the inverse of the state's lifetime. At the same 
time, the continua are rotated by an angle of $2\theta$ into the lower-half 
complex plane.

If the Hamiltonian is represented exactly, only the energies of the continua 
and the resonances embedded therein are affected by CS, while bound states 
have Im($E$) = 0 even though their wave functions change. Also, the complex 
eigenvalues of the resonances are independent of $\theta$ if it is larger 
than the critical value\cite{balslev71}
\begin{equation} \label{eq:cs3}
\theta_\text{c} = 1/2 \, \text{arctan} [\Gamma/2 \, (E_\text{R}-E_\text{t})]
\end{equation}
with $E_\text{t}$ as threshold energy. Above the same critical angle, the resonance
wave functions are $L^2$ integrable and thus amenable to a treatment with bound-state
methods. In the context of Auger decay, Eq. \eqref{eq:cs3} implies that very small 
scaling angles are sufficient to uncover the resonances and make their wave functions
$L^2$ integrable. If we consider for a back-of-the-envelope estimate the core-ionized
state of neon ($\Gamma \approx 0.25$ eV, $E_\text{R}-E_\text{t} \approx 800$ eV), a 
scaling angle of less than 0.01$^\circ$ should be sufficient. This is in contrast to,
for example, the temporary anion $\text{N}_2^-$ ($\Gamma \approx 0.4$ eV, 
$E_\text{R}-E_\text{t} \approx 2.3$ eV) \cite{jagau17} where the critical 
angle is ca. 5$^\circ$. Core-ionized wave functions are thus on the verge 
of $L^2$ integrability, which distinguishes them from other types of resonances. 
In actual calculations with a finite basis, $E_\text{res}$ does depend on 
$\theta$; the optimal value is usually found through minimizing $|\text{d}E/
\text{d}\theta|$.\cite{moiseyev78,jagau17} For this purpose, trajectories 
$E(\theta)$ need to be calculated, which is the main reason that complex-scaled 
methods are more computationally expensive than their real-valued counterparts. 

CS has the major disadvantage that it cannot be applied to molecules because the 
complex-scaled electron-nuclear attraction is not dilation analytic within the 
Born-Oppenheimer approximation.\cite{moiseyev11} A possible solution is exterior 
scaling,\cite{simon79} where the area close to the nuclei is not scaled. In the
context of Gaussian basis sets, this can be realized by the method of complex 
basis functions (CBFs),\cite{mccurdy78,moiseyev79} which relies on the identity 
\begin{equation} \label{eq:cs4}
\frac{\langle \Psi(r) | \Hat{H}(r\text{e}^{\text{i}\theta}) | \Psi(r) \rangle}
{\langle \Psi(r) | \Psi(r) \rangle} = \frac{ \langle 
\Psi(r\text{e}^{-\text{i}\theta}) | \Hat{H}(r) | \Psi(r\text{e}^{-\text{i}\theta}) \rangle}
{\langle \Psi(r\text{e}^{-\text{i}\theta}) | \Psi(r\text{e}^{-\text{i}\theta}) \rangle} 
\end{equation}
and the fact that scaling the coordinates of the basis functions according 
to the right-hand side of Eq. \eqref{eq:cs4} is equivalent to scaling their 
exponents in the same way. Since it is possible to scale only selected 
basis functions ---in the computational practice hitherto the most diffuse 
shells\cite{white15a}--- dilation analyticity is preserved and CBF methods 
are applicable to molecules.

A further advantage of CBF methods over CS is that changes in the bound-state 
and resonance wave functions stemming from Eq. \eqref{eq:cs1} are smaller. 
As a result, Im($E)$ of bound states, which is zero in the full basis-set 
limit, is smaller by orders of magnitude in CBF calculations than in CS 
calculations. 

We add here that the scaling angle can be chosen to be complex-valued in CBF 
and CS methods, that is, $\text{e}^{-\text{i}\theta} = \alpha\cdot
\text{e}^{-\text{i}\theta_\text{R}}$; $\alpha, \theta_\text{R} \in \mathbb{R}$.
\cite{moiseyev79,moiseyev11} The factor $\alpha$ represents an optimization of 
the exponents of the basis functions and is related to the stabilization 
method where resonances are identified from changes in the energy upon scaling 
the exponents.\cite{hazi70,hazi76} 

CBF methods offer access to different types of molecular electronic resonances 
as illustrated by many recent applications.\cite{white15a,white15b,white17,
jagau18,hernandez19,hernandez20} However, no work on molecular Auger decay 
has been reported. As we will show in Sec. \ref{sec:atres}, a straightforward 
application of computational protocols developed for other resonances results 
in zero decay widths, that is, such CBF calculations are blind to Auger decay. 
Several changes are necessary to uncover the decaying character of core-ionized 
states. 

%%%%%%%%%%%%%%%%%%%%%%%%%%%%%%%%%%%%%%%%%%%%%%%%%%%%%%%%%%%%%%%%%%%%%%%%%%%%%%

\subsection{Complex-variable coupled-cluster methods}
\label{sec:thcc}
In CC theory, the wave function is obtained from the Hartree-Fock (HF) state 
$| \Psi_0 \rangle$ by the action of the cluster operator $\Hat{T}$ according to  
\begin{equation} \label{eq:cc1}
|\Psi_\text{CC} \rangle = \text{e}^{\Hat{T}}|\Psi_0 \rangle = 
(1 + \Hat{T} + \Hat{T}^2/2! + \Hat{T}^3/3! + \dots)|\Psi_0 \rangle~.
\end{equation}
In CBF-CC methods, $| \Psi_0 \rangle$ is always complex-valued while different 
approaches are possible for CS-CC methods.\cite{bravaya13,jagau17} In this work, 
all CS calculations are based on a CS-HF reference. 

Inclusion of different excitation levels in $\Hat{T}$ gives rise to a hierarchy 
of methods that converges smoothly to the exact solution. Here, we use CCSD 
where $\Hat{T} = \Hat{T}_1 + \Hat{T}_2$. The exponential parametrization in Eq.
\eqref{eq:cc1} ensures size-extensivity and inclusion of the most relevant
higher excitations through products of $T_1$ and $T_2$.~\cite{shavitt09}

Inserting Eq. \eqref{eq:cc1} into the Schr\"odinger equation, one obtains
\begin{equation} \label{eq:cc2}
\Hat{H} \text{e}^{\Hat{T}} | \Psi_0 \rangle = 
E \text{e}^{\Hat{T}} | \Psi_0 \rangle \quad \Leftrightarrow \quad
\text{e}^{-\Hat{T}} \Hat{H} \text{e}^{\Hat{T}} | \Psi_0\rangle = 
E| \Psi_0\rangle, 
\end{equation}
where $\text{e}^{-\Hat{T}}\Hat{H}\text{e}^{\Hat{T}} = \bar{H}$ is the 
similarity-transformed Hamiltonian. Projection of Eq. \eqref{eq:cc2} onto 
the HF determinant and the singly and doubly excited determinants determines 
the CCSD energy and amplitudes, respectively. 

To describe core-ionized states subject to Auger decay within CC theory, we 
employ two computational strategies as illustrated in Fig. \ref{fig:eomccsd}: 
In the $\Delta$CCSD approach, one performs two separate CCSD calculations 
based on HF determinants for the neutral state and the core-ionized state. 
In this work, the latter state is always described in a spin-unrestricted 
manner. We determine the optimal scaling angle from the difference of the 
two CCSD energies; in accordance with previous reports\cite{white17} we 
find that this approach usually leads to much smaller values for 
$|\text{d}E/\text{d}\theta|$ than determining $\theta$ from the 
total energy. If not specified otherwise, we recomputed all energies in the 
0--45$^\circ$ range in steps of 1$^\circ$. The total decay width $\Gamma$ is 
then evaluated from the difference of the two imaginary energies. We reiterate 
that Im($E$) of the neutral state would be zero in exact theory, but has a 
significant value especially in CS-based calculations (see Sec. \ref{sec:thcs}). 

The other method we use is EOMIP-CCSD.\cite{nooijen93,stanton94} Based on a 
CCSD wave function for a neutral molecule, biorthogonal right and left wave 
functions for a core-ionized state are constructed in EOMIP-CCSD as 
\begin{align}\label{eq:eomwf}
|\Psi_\text{EOMCC} \rangle &= \Hat{R} 
\text{e}^{\Hat{T}} | \Psi_0 \rangle~, \\
\langle\Psi_\text{EOMCC}| &= \langle \Psi_0 | \Hat{L}^\dagger \text{e}^{-\Hat{T}}~.
\end{align}
Here, the excitation operators $\Hat{R}$ and $\Hat{L}$ are truncated at the same 
level as $\Hat{T}$, meaning that they include 1-hole ($1\text{h}$) and 
2-hole-1-particle ($2\text{h}1\text{p}$) excitations in EOMIP-CCSD. Insertion 
of Eq. \eqref{eq:eomwf} into the Schr\"odinger equation and projection onto the 
$1\text{h}$ and $2\text{h}1\text{p}$ excitation manifolds results in an 
eigenvalue equation for $\bar{H}$. The total decay width $\Gamma$ is directly 
obtained from the imaginary part of the eigenvalues according to 
Eq.~\eqref{eq:cs2}. The optimal scaling angle is also determined from these 
eigenvalues of the EOMIP-CCSD equations.

The computational cost of both approaches scales as $N^6$ 
with system size. However, while $\Delta$CCSD involves two separate CCSD 
calculations with $N_\text{occ}^2 N_\text{virt}^4$ cost, one needs to carry 
out only one such calculation for the reference state in the EOMIP-CCSD 
approach; the EOMIP step itself scales as $N_\text{occ}^2 N_\text{virt}^3$. 
As a consequence, the EOMIP-CCSD approach typically entails lower computational 
cost.

All calculations presented in this article were carried out with the
complex-variable CCSD and EOM-CCSD codes~\cite{bravaya13,zuev14} implemented 
in the Q-Chem software.\cite{epifanovsky21} 

%Both $\Delta$CCSD and EOMIP-CCSD scale identically with the
%system size at $N^6$, i.\,e. $N_\text{occ}^2 N_\text{virt}^4$. However, the
%EOMIP-CCSD step can benefit from the use of the Davidson algorithm, often
%resulting in a lower computational cost. Since the EOMIP-CCSD calculation also
%requires a CCSD solution for the reference state, the computational cost has the
%same order of magnitude for both methods but is typically more favorable for
%EOMIP-CCSD calculations.
%Both approaches require an initial CCSD solution for the neutral molecule. In the $\Delta$CCSD method, a second CCSD calculation is then performed for the core-ionized state, while EOMIP-CCSD requires solving for the amplitudes in $\Hat{R}$ while retaining the $\Hat{T}$ operator used for the reference calculation.

%%%%%%%%%%%%%%%%%%%%%%%%%%%%%%%%%%%%%%%%%%%%%%%%%%%%%%%%%%%%%%%%%%%%%%%%%%%%%%%%%%

\subsection{Structure of the core-ionized wave function} 
\label{sec:thst}

\begin{figure*} \centering
\includegraphics[width=.91\linewidth]{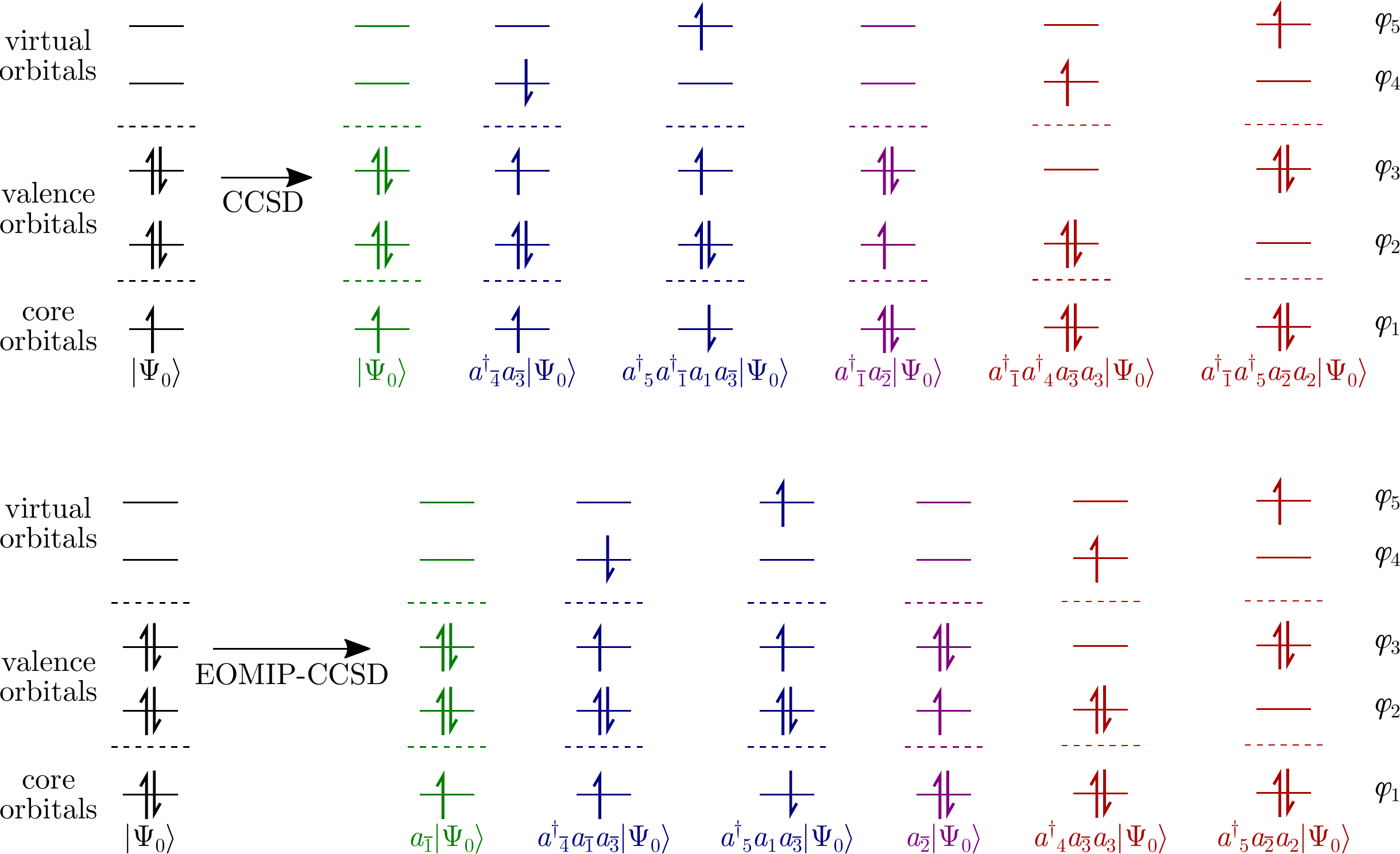}
\caption{Determinants included in CCSD (upper part) and EOMIP-CCSD (lower part) 
wave functions for a core-ionized state in a Hilbert space spanned by 5 orbitals. 
See text for further explanation.}
\label{fig:eomccsd}
\end{figure*}

Fig. \ref{fig:eomccsd} displays the structures of the complex-scaled CCSD and 
EOMIP-CCSD wave functions of a core-ionized state, which shows that both 
methods are capable of describing Auger decay: this is primarily achieved 
by means of the doubly-excited determinants marked in red, where the core 
hole has been filled with an electron, while a second valence electron has
been excited into the virtual space. 

Our numerical results (see Sec. \ref{sec:atres}) confirm that the red 
configurations in Fig. \ref{fig:eomccsd} are almost exclusively responsible 
for the decay width in the $\Delta$CCSD approach as one would expect. This 
is, however, not the case in EOMIP-CCSD as we will discuss also in Sec. 
\ref{sec:atres}. In both approaches, the red determinants are of very minor 
relevance for the real part of the energy and their amplitudes are typically 
orders of magnitude smaller than those of the blue configurations. This 
corroborates the validity of CVS methods since the CVS projector precisely 
removes the red determinants from the wave function.\cite{cederbaum80} 

The green determinants represent zeroth-order descriptions of the core-vacant 
state and, while they carry by far the largest weight in the wave functions, 
are not relevant to our further analysis. The blue determinants comprise single 
and double excitations and represent orbital relaxation as well electron 
correlation effects. They play different roles in $\Delta$CCSD and EOMIP-CCSD:
In the latter approach, the HF wave function $| \Psi_0 \rangle$ is optimized 
for the neutral ground state and electron correlation is subsequently treated 
for this state as well. The blue determinants are thus indispensable in 
EOMIP-CCSD to model the substantial relaxation in the charge distribution 
due to the core hole as well as differential electron correlation. In 
the former approach, i.\,e. $\Delta$CCSD, relaxation effects are already 
contained in $| \Psi_0 \rangle$ through changed orbital shapes and energies. 
The doubly-excited blue determinants thus describe primarily electron 
correlation and their singly-excited counterparts secondary relaxation 
effects resulting therefrom.

%%%%%%%%%%%%%%%%%%%%%%%%%%%%%%%%%%%%%%%%%%%%%%%%%%%%%%%%%%%%%%%%%%%%%%%%%%%%

\subsection{Energy decomposition analysis and partial decay widths}
\label{sec:thpw}

To substantiate the qualitative discussion from the preceding section, we use 
energy decomposition analysis. This allows us to identify contributions from 
individual excitations to the imaginary energy, that is, the total decay 
width. In addition, we get access to partial widths corresponding to decay 
into particular channels. 

For $\Delta$CCSD, we use two approaches. We either decompose directly the 
CCSD energy 
\begin{equation} \label{eq:decomp1b}
E = E_\text{HF} + \sum_{ijab} \Big( \frac{1}{4} t_{ij}^{ab} 
+ \frac{1}{2} t_i^a t_j^b \Big) \; \langle ij || ab \rangle
\end{equation}
or, alternatively, use an expression in terms of the reduced one-electron 
and two-electron CCSD density matrices $\mathbf{D}^\text{CCSD}$ 
and $\boldsymbol{\Gamma}^\text{CCSD}$. The latter reads 
\begin{align} \label{eq:decomp1}
E &= \langle \Psi_0 \, | \, (1 + \Hat{\Lambda}) \, \text{e}^{-\Hat{T}} \Hat{H} 
\text{e}^{\Hat{T}} \, | \, \Psi_0 \rangle \nonumber \\
&= E_\text{HF} + \sum_{pq} \, f_{pq}  \, \langle \Psi_0 | (1 + \Hat{\Lambda}) 
\text{e}^{-\Hat{T}} \{ p^\dagger q \} \text{e}^{\Hat{T}} | \Psi_0 \rangle 
\nonumber \\
& \quad + \frac{1}{4} \sum_{pqrs} \langle pq||rs \rangle \, \langle \Psi_0 | 
(1 + \Hat{\Lambda}) \text{e}^{-\Hat{T}} \{p^\dagger q^\dagger sr \} 
\text{e}^{\Hat{T}} | \Psi_0 \rangle \nonumber \\
&= \, E_\text{HF} \, + \, \sum_{pq} D^\text{CCSD}_{pq} \, f_{pq} \, 
+ \, \frac{1}{4} \, \sum_{pqrs} \, \Gamma^\text{CCSD}_{pqrs} \, 
\langle pq||rs \rangle
\end{align}
with $\Hat{\Lambda}$ as the well-known deexcitation operator from CC gradient
theory \cite{shavitt09} and $f_{pq}$ and $\langle pq||rs\rangle$ as elements 
of the Fock matrix and antisymmetrized two-electron integrals, respectively.

For EOMIP-CCSD, the corresponding expression reads
\begin{align} \label{eq:decomp2}
E &= \langle \Psi_0 \, | \, \hat{L}^\dagger \, \text{e}^{-\Hat{T}} \Hat{H} 
\text{e}^{\Hat{T}} \, \hat{R} \, | \, \Psi_0 \rangle \nonumber \\
&= E_\text{HF} + \sum_{pq} \, f_{pq} \, \langle \Psi_0 | \hat{L}^\dagger \, 
\text{e}^{-\Hat{T}} \{ p^\dagger q \} \text{e}^{\Hat{T}} \, \hat{R} | 
\Psi_0 \rangle \nonumber \\
& \quad + \frac{1}{4} \sum_{pqrs} \langle pq||rs \rangle \, \langle \Psi_0 | 
\hat{L}^\dagger \, \text{e}^{-\Hat{T}} \{p^\dagger q^\dagger sr \} 
\text{e}^{\Hat{T}} \, \hat{R} | \Psi_0 \rangle \nonumber \\
&= \, E_\text{HF} \, + \, \sum_{pq} D^\text{EOMIP}_{pq} \, f_{pq} \, 
+ \, \frac{1}{4} \, \sum_{pqrs} \, \Gamma^\text{EOMIP}_{pqrs} \, 
\langle pq||rs \rangle
\end{align}
and differs from Eq. \eqref{eq:decomp1} thus only in the definition of the 
density matrices $\mathbf{D}^\text{EOMIP}$ and $\boldsymbol{\Gamma}^\text{EOMIP}$.

To compute partial widths from Eq. \eqref{eq:decomp1b}, we use a modified 
$\Hat{T}_2$ operator, where amplitudes corresponding to a particular 
decay channel have been set to zero. For the corresponding decomposition
based on Eqs. \eqref{eq:decomp1} and \eqref{eq:decomp2}, we use modified 
density matrices: After convergence of the CCSD or EOMIP-CCSD equations, 
we set to zero the amplitudes in $\Hat{T}$ and $\Hat{\Lambda}$ or $\Hat{R}$ 
and $\Hat{L}$ that correspond to a particular decay channel. For EOMIP-CCSD, 
this is done such that spin-completeness is preserved. We note that decay 
into a particular channel is usually represented by multiple excitations 
that differ in the energy of the virtual orbital (see Fig. \ref{fig:eomccsd}). 
Using these approaches, we can disable Auger decay channel by channel 
until we arrive at versions of $\Hat{T}$, $\Hat{\Lambda}$, $\Hat{R}$, 
and $\Hat{L}$ that are used in CVS-CCSD and CVS-EOMIP-CCSD. When evaluated 
from these CVS-like operators, Eqs. \eqref{eq:decomp1b}--\eqref{eq:decomp2} 
yield zero decay widths. 

%%%%%%%%%%%%%%%%%%%%%%%%%%%%%%%%%%%%%%%%%%%%%%%%%%%%%%%%%%%%%%%%%%%%%%%%%%%%%%

\section{Numerical Analysis of Auger Decay of $\text{Ne}^+$ 
($1\text{s}^{-1}$)} \label{sec:atres}
As alluded to in Sec. \ref{sec:thcs}, the application of computational 
protocols developed for other types of resonances works well for CS but 
not for CBF methods. To analyze this further, we use the 1s$^{-1}$ state 
of the neon atom as a test case. This system is a simple and frequently 
studied example of Auger decay.\cite{skomorowski21b,howat77,albiez90,
coreno99,muller17} There are five main decay channels leading to the
$^1$D ($2\text{p}^{-2}$), $^1$S ($2\text{p}^{-2}$), $^3$P 
($2\text{s}^{-1}2\text{p}^{-1}$), $^1$P ($2\text{s}^{-1}2\text{p}^{-1}$), 
and $^1$S ($2\text{s}^{-2}$) states of Ne$^{2+}$. 

\subsection{Total Auger decay width from complex scaling} \label{sec:netot}

\begin{table}
% compared to experimental and theoretical numbers taken from the literature.
\caption{Energies and half-widths of Ne$^+$ (1$\text{s}^{-1}$) computed with 
CS-EOMIP-CCSD and CS-$\Delta$CCSD using different basis sets. The basis-set 
suffix +3s3p refers to the addition of 3 diffuse s and p shells with an 
even-tempered exponent spacing of 2. Energies in eV. Half-widths
in meV. Results for additional basis sets can be found in the SI.}

\begin{ruledtabular} \begin{tabular}{llcrr}
Method & Basis set & $\theta_\text{opt} / ^\circ$ & $\text{Re}(E)$ & 
$\text{Im}(E)$ \\ \hline
EOMIP-CCSD & aug-cc-pCVTZ+6s6p6d & \multicolumn{3}{c}{No minimum in
$|\text{d}E/\text{d}\theta|$} \\
EOMIP-CCSD & cc-pCVQZ & \multicolumn{3}{c}{No minimum in
$|\text{d}E/\text{d}\theta|$} \\
EOMIP-CCSD & aug-cc-pCVQZ & 10 & 871.26 & $-73$ \\
EOMIP-CCSD & cc-pCV5Z & 11 & 871.22 & $-113$ \\
EOMIP-CCSD & aug-cc-pCV5Z & 13 & 871.20 & $-114$ \\
EOMIP-CCSD & aug-cc-pCV5Z+3s3p & 14 & 871.20 & $-109$ \\
EOMIP-CCSD & cc-pCV6Z & 12 & 871.22 & $-97$ \\ \hline
$\Delta$CCSD & aug-cc-pCV5Z & 12 & 869.53 & $-104$ \\
$\Delta$CCSD & aug-cc-pCV5Z+3s3p&13 & 869.53 & $-101$ \\ \hline
Fano & \multicolumn{2}{c}{} & 
870.12$^\text{a}$ & $-109^\text{b}$ \\
Experiment & --- & --- & 870.17$^\text{a}$ & $-129^\text{c}$ \\
\end{tabular} \end{ruledtabular} \label{tab:cs}
\footnotetext{From Ref. \citenum{coreno99}, theoretical value computed 
using CI wave functions.\cite{colle93}}
\footnotetext{From Ref. \citenum{skomorowski21b}, computed using Fano's 
theory based on EOMIP-CCSD and EOMDIP-CCSD wave functions.}
\footnotetext{From Ref. \citenum{muller17}.}
\end{table}

Tab. \ref{tab:cs} shows core-ionization energies and Auger decay widths 
of Ne$^+$ computed with CS-EOMIP-CCSD and CS-$\Delta$CCSD. This confirms 
the conclusion from Sec. \ref{sec:thst} that both methods are able to 
describe Auger decay. $\Delta$CCSD yields a somewhat more accurate result 
for the ionization energy as compared to EOMIP-CCSD, which is in line with 
previous findings using real-valued CC methods.\cite{zheng19,matthews20} 
The decay widths differ by less than 10\,\% and are well in line with 
earlier theoretical results.\cite{skomorowski21b} The 
underestimation of the decay width by 10-20\,\% as compared to the 
experimental value\cite{muller15} can be related to the fact that double 
Auger decay and other processes involving more than two electrons are not 
described within the CCSD approximation. The corresponding decay channels 
are not present in our calculations.

%Parts of the differences can be explained by
%insufficiencies of the correlation method: three- or more-electron decay
%processes such as double Auger decay are not described by the
%Coupled-Cluster wavefunction since only disconnected triple and higher excitations
%are present. The exclusion of these decay channels explains why all calculated
%total widths are lower than the experimental value. The similarity of our results
%with the results of the application of Fano's theory to EOM-CCSD wave functions
%supports this assertion.}

Tab. \ref{tab:cs} also illustrates that extra diffuse shells are not needed 
to describe Auger decay of Ne$^+$. This is unlike to low-lying temporary 
anions and Stark resonances formed in static electric fields, where these 
extra shells are vital to obtain accurate decay widths with CS methods.\cite{
white15a,white17,hernandez20} On the other hand, requirements towards the 
valence part of the basis set are as high as in CS calculations of other 
types of resonances. It appears that aug-cc-pCVQZ is the smallest basis 
set that is able to capture the decaying character of the wave function 
and even this basis set recovers only 2/3 of the decay width computed with 
aug-cc-pCV5Z. 

%%%%%%%%%%%%%%%%%%%%%%%%%%%%%%%%%%%%%%%%%%%%%%%%%%%%%%%%%%%%%%%%%%%%%%%%%%%%

\subsection{Partial Auger decay widths from complex scaling} \label{sec:nepw}

To compute partial decay widths, we decomposed the total CS-CCSD decay 
width of Ne$^+$ ($1\text{s}^{-1}$) on the basis of Eqs. \eqref{eq:decomp1b} 
and \eqref{eq:decomp1} by setting to zero amplitudes in $\Hat{T}_2$ and 
$\Hat{\Lambda}_2$ that create the ``red'' determinants in Fig. \ref{fig:eomccsd}. 
Our results are compiled in Tab. \ref{tab:cspw} and compared with results 
obtained using a combination of Fano's approach with EOM-CC theory\cite{
skomorowski21b} as well as with experimental values.\cite{muller17,albiez90} 
As branching ratios and total decay widths are usually determined by separate 
experiments, we derived the experimental values for the partial widths in Tab. 
\ref{tab:cspw} by multiplying branching ratios from Ref. \citenum{albiez90} 
with the result of a recent measurement of the total decay width.\cite{muller17}

\begin{table}
\caption{Partial decay half-widths for the 5 decay channels of Ne$^+$ 
($1\text{s}^{-1}$). All values in meV.}
\begin{ruledtabular}\begin{tabular}{lrrrrr}
Decay channel & CS/$\Delta^\text{a}$ & CS/$\Delta^\text{b}$ & CS/EOM$^\text{c}$ &
Fano$^\text{d}$ & Experiment$^\text{e}$ \\ \hline
all & 122.3 & 133.7 & 210.9 & 109.1 & 128.5(30) \\
$^1$D (2p$^{-2}$) & 74.7 & 81.5 & 133.4 & 58.8 & 78.2(21) \\
$^1$P (2s$^{-1}$2p$^{-1}$) & 27.7 & 29.3 & 21.6 & 19.6 & 22.1(7) \\
$^3$P (2s$^{-1}$2p$^{-1}$) & 6.6 & 6.6 & 43.2 & 11.9 & 8.1(3) \\
$^1$S (2s$^{-2}$) & 9.3 & 8.8 & 1.0 & 13.6 & 7.9(3) \\
$^1$S (2p$^{-2}$) & 7.0 & 7.6 & 12.8 & 5.3 & 12.2(4) \\
\end{tabular} \end{ruledtabular} \label{tab:cspw}
\footnotetext{This work, computed using CS-$\Delta$CCSD/aug-cc-pCV5Z and 
Eq. \eqref{eq:decomp1}.}
\footnotetext{This work, computed using CS-$\Delta$CCSD/aug-cc-pCV5Z and 
Eq. \eqref{eq:decomp1b}.}
\footnotetext{This work, computed using CS-EOMIP-CCSD/aug-cc-pCV5Z and 
Eq. \eqref{eq:decomp2}.}
\footnotetext{From Ref. \citenum{skomorowski21b}, computed using Fano's 
theory based on EOMIP-CCSD and EOMDIP-CCSD wave functions.}
\footnotetext{From Refs. \citenum{muller17,albiez90}.}
\end{table}

Tab. \ref{tab:cspw} illustrates overall excellent agreement between our 
CS-CCSD partial widths and those from experiment; the experimental values 
are reproduced with a root mean square deviation of 4~meV.
However, there are several issues that deserve a discussion: 

First, the half-widths computed with Eqs. \eqref{eq:decomp1b} and 
\eqref{eq:decomp1} are not identical. This is due to the structure 
of the CCSD density matrices in Eq. \eqref{eq:decomp1}: A term such as 
$\Hat{\Lambda}_{\text{d-decay}} \cdot \Hat{T}_{\text{s-decay}}$ is set 
to zero when computing the partial widths for either decay channel and 
thus counted twice. This also causes that the sum of the partial 
half-widths of the 5 channels (125 meV) is not identical to the total 
half-width in Tab. \ref{tab:cspw} (122 meV). On the other hand, no 
double counting occurs in Eq. \eqref{eq:decomp1b} and the corresponding 
partial widths are strictly additive.

A second observation is that neither Eq. \eqref{eq:decomp1b} nor Eq. 
\eqref{eq:decomp1} yield half-widths that sum up to the total CS-CCSD 
half-width reported in Tab. \ref{tab:cs}. When removing all determinants 
marked in red in Fig. \ref{fig:eomccsd}, we obtain values of 134 and 
122 meV from Eqs. \eqref{eq:decomp1b} and \eqref{eq:decomp1}, respectively, 
whereas the value from Tab. \ref{tab:cs} is 104 meV. 

This discrepancy stems from three origins: First, $\Gamma/2$ in Tab.~\ref{tab:cs}
is evaluated from the energy difference between the core-ionized 
and the neutral ground state, whereas Eqs. \eqref{eq:decomp1b} and 
\eqref{eq:decomp1} are applied only to the core-ionized state. Second, 
the ``red'' determinants in Fig. \ref{fig:eomccsd} contribute to $\Gamma/2$ 
not only through $\Hat{T}_2$ but also through $\Hat{T}_1 \cdot \Hat{T}_1$. 
Third, the determinants marked in blue and purple in Fig. \ref{fig:eomccsd} 
deliver a non-negligible contribution to $\Gamma/2$, but their assignment 
to a particular decay channel is not straightforward. We neglected these 
contributions in the values reported in Tab. \ref{tab:cspw} but we note 
that the ``blue'' determinants are related to shake-up and shake-off 
processes, which are well known in the context of interchannel coupling 
in Auger decay.\cite{howat77,colle90}

An equivalent decomposition of the EOMIP-CCSD decay width was performed 
on the basis of Eq. \eqref{eq:decomp2} by setting to zero elements of 
$\Hat{R}_2$ and $\Hat{L}_2$. These results are also contained in Tab. 
\ref{tab:cspw}. It is apparent that the EOMIP-CCSD partial widths are 
very unreliable: By removing all excitations into the ``red'' determinants 
from $\Hat{R}_2$ and $\Hat{L}_2$, we obtain for $\Gamma/2$ a value of 
211 meV from Eq. \eqref{eq:decomp2}, whereas the imaginary part of the 
eigenvalue of the EOMIP-CCSD equations is 114 meV (see Tab. \ref{tab:cs}). 

The reasons for this failure are similar in origin to the much smaller 
discrepancies between the CCSD values discussed before. First, Eq. 
\eqref{eq:decomp2} is an expression for $E-E_\text{HF}$, while the 
EOMIP-CCSD equations yield $E-E_\text{CCSD}$ as eigenvalue. Albeit 
zero in exact theory, the imaginary energy of the neutral CCSD reference 
state amounts to $\sim 1.3\cdot10^{-3}$ a.u. or 35 meV in our
calculations. Second, the ``blue'' and ``purple'' determinants from Fig.
\ref{fig:eomccsd} are again neglected. Third, the ``red'' determinants in Fig.
\ref{fig:eomccsd} cannot only be created by $\Hat{R}_2$ but also by combinations
of $\Hat{R}_1$ and $\Hat{T}_2$ as well as $\Hat{R}_1$ and $\Hat{T}_1$. These
contributions to $\Gamma/2$ are substantial and their neglect is the reason that
we observe much larger discrepancies between Tabs. \ref{tab:cs} and \ref{tab:cspw}
for EOMIP-CCSD than for $\Delta$CCSD. 

%FOLLOWING UNCLEAR TO ME, BLUE DETS WILL ALWAYS HAVE GAMMA.NEQ.0
%In the physical system, there is no decay resulting from these channels and we can rationalize their contributions to the decay width by the strong perturbation to the wave function introduced by scaling the full Hamiltonian, including matrix elements involving core orbitals. 

%%%%%%%%%%%%%%%%%%%%%%%%%%%%%%%%%%%%%%%%%%%%%%%%%%%%%%%%%%%%%%%%%%%%%%%%%%%

\subsection{Analysis of the orbital basis} \label{sec:neorb}

To understand the basis-set dependence documented in Tab.~\ref{tab:cs}, we 
decomposed the CS-CCSD partial decay width of the $^1$D(2p$^{-2}$) channel
further into contributions from different excitations. The 8 d shells in the 
aug-cc-pCV5Z basis for Ne give rise to 8 sets of virtual orbitals with 
d-symmetry. We computed their contributions to $\Gamma/2$ by setting to zero 
in Eq.~\eqref{eq:decomp1} those amplitudes $t_{ij}^{ab}$ and $\lambda_{ab}^{ij}$  
where $i=j=2\text{p}$ and $a=1\text{s}$, $b=n\text{d}$ or $a=n\text{d}$,
$b=1\text{s}$.
The results are presented in Tab. \ref{tab:orbs1} together with the corresponding 
orbital energies. 

\begin{table}
\caption{Contributions $\Delta$Im($E$) of different d orbitals ($n=3-10$) 
in meV to the CS-CCSD partial half width of the $^1$D(2p$^{-2}$) channel of 
Ne$^+$($1\text{s}^{-1}$) computed with the aug-pCV5Z basis set at
$\theta_\text{opt} = 12^\circ$. All orbital energies $\varepsilon$ in a.u. and the
three-particle overlaps $O$ (in \AA$^3$ Bohr$^{-4.5}$) of the d orbitals with the
2p orbitals are also given. See text for further details.}
\begin{ruledtabular}\begin{tabular}{lrrrrr}
$n$ & $\Delta$Im($E$) & Re($O$) & Im($O$) & Re($\varepsilon$) & 
Im($\varepsilon$) \\ \hline
3 & $-$0.1 & $-0.9$ & 2.8 $\cdot 10^{-8}$ & 0.60 & $-$0.27 \\
4 & $-$1.6 & 3.5 & $-$7.3 $\cdot 10^{-8}$ & 1.88 & $-$0.79 \\
5 & $-$6.5 & 7.4 & 1.4 $\cdot 10^{-7}$ & 4.79 & $-$2.12 \\
6 & $-$18.0 & 8.8 & 5.8 $\cdot 10^{-7}$ & 12.11 & $-$5.74 \\
7 & $-$46.7 & 5.8 & 7.2 $\cdot 10^{-7}$ & 31.61 & $-$15.81 \\
8 & $-$2.6 & $-$1.9 & $-$4.5 $\cdot 10^{-7}$ & 88.28 & $-$45.27 \\
9 & $-$0.1 & 0.4 & 9.6 $\cdot 10^{-8}$ & 259.60 & $-$132.53 \\
10 & $-$0.0 & 0.1 & 1.5 $\cdot 10^{-8}$ & 791.25 & $-$399.75 \\
\end{tabular} \end{ruledtabular} \label{tab:orbs1}
\end{table}

Tab. \ref{tab:orbs1} illustrates that the $^1$D partial decay width arises 
almost exclusively from excitations into determinants in which the 5d, 6d, 
and 7d orbitals are occupied, while excitations into the remaining d orbitals 
contribute only 6\,\%. We note that for the 7d orbital, which delivers the 
largest contribution to $\Gamma/2$, Re($\varepsilon$) $\approx$ 860 eV, 
which is close to the kinetic energy of the emitted Auger electron (804 
eV).\cite{albiez90} This observation offers an explanation why 
scaling of diffuse basis functions is necessary for the 
description of low-lying temporary anions, where the outgoing electron 
has just a few eV, while functions with larger exponent need to be scaled 
to describe Auger decay. This is the reason why calculations in which only 
diffuse functions are scaled are blind to Auger decay.

\begin{figure}
\includegraphics[width=\linewidth]{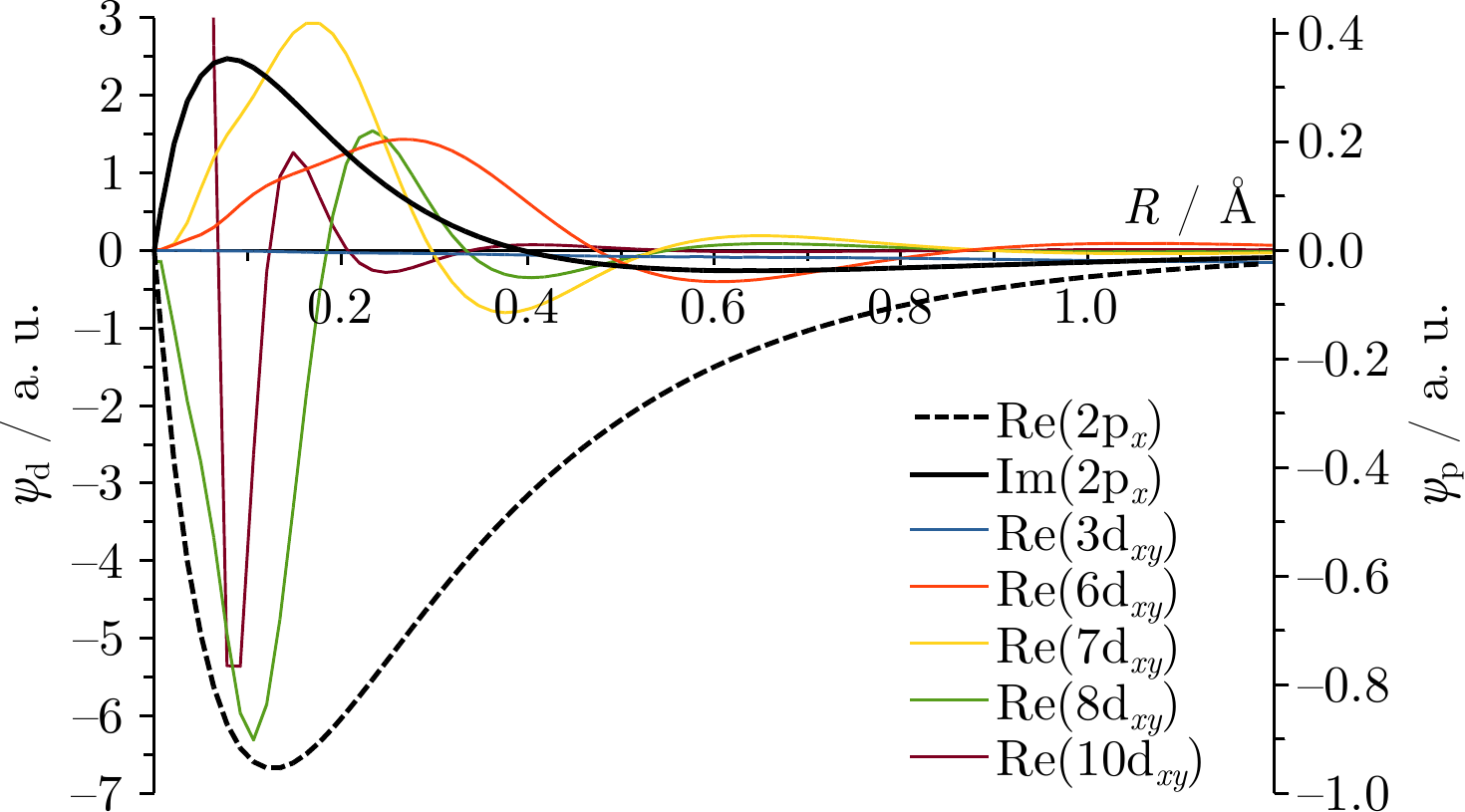}
\caption{Plots of the 2p$_x^\upbeta$ and selected d$_{xy}^\upalpha$ orbitals 
along the $xy$ diagonal obtained from a CS-UHF calculation for 
Ne$^+$($1\text{s}_\upalpha^{-1}$) using the aug-cc-pCV5Z basis set.}
\label{fig:orbitals} 
\end{figure}

To analyze this further, we computed the values of the d$_{xy}^\upalpha$ 
orbitals along the $xy$ diagonal at $z=0$ and compared them to those of 
the 2p$_x^\upalpha$ and 2p$_x^\upbeta$ orbitals. Along this line, the 
aforementioned orbitals have no nodal plane due to the angular part that
would complicate the analysis. Plots of the real and imaginary parts of
the 2p$_x^\upbeta$ orbitals and a few selected d$_{xy}^\upalpha$ orbitals
are presented in Fig.~\ref{fig:orbitals}. As shown in the SI, the differences
between the 2p$_x^\upalpha$ and 2p$_x^\upbeta$ orbitals are small and not 
relevant for the further discussion. 

The d orbital with the lowest energies (e.\,g. 3d$_{xy}$) are very diffuse
and reach their maximum amplitude at a distance of 1.3~\AA\ from the nucleus, 
where the amplitude of the 2p$_x$ orbital is negligible. On the other hand, 
the d orbitals with high energies (e.\,g. 10d$_{xy}$) have 
strong oscillations close to the nucleus. Only the d orbitals in between 
(e.\,g. 6d$_{xy}$, 7d$_{xy}$), which together yield more than 90~\% of 
the decay width, have large amplitudes close to the nucleus and at the same 
time no node less than 0.3~\AA\ away from it so that their overlap with the 
2p$_x$ orbital is large. 

We quantify this overlap by integrating over the radial coordinate of the product 
of three orbitals as follows:
\begin{equation} \label{eq:orb1}
O = \int_0^\infty \text{d}r \; 4\uppi r^2 \;
\varphi_{\text{2p},x}^{\upalpha}(r) \; 
\varphi_{\text{2p},y}^{\upbeta}(r) \; \varphi_{n\text{d}, xy}^\upalpha(r) ~.
\end{equation}
Eq. \eqref{eq:orb1} is formulated for decay into the $^1$D state of Ne$^{2+}$ 
but can be easily generalized to other systems. We note that, because p orbitals 
are \textit{ungerade} and d orbitals are \textit{gerade}, only the three-orbital 
overlap defined according to Eq. \eqref{eq:orb1} is nonzero, whereas the integral 
over a simple product of 2p$_x$ and $n$d$_{xy}$ along the $xy$ diagonal would
vanish.
%results from numerical integration of |Re(4\uppi r^2 \;
%\varphi_{\text{2p,y}}^{\upalpha}(r) \; \varphi_{\text{nd, xy}}^\upalpha(r))| are
%between 7e-13 and 5e-12

The values for $O$ calculated from Eq. \eqref{eq:orb1} are listed in
Tab.~\ref{tab:orbs1}.
This shows that Re($O$) and Im($O$) are indicators of the 
contribution of a particular d orbital to the decay width; the d orbitals 
with the largest overlaps with 2p$_x$ also contribute most to $\Gamma/2$. 
The description of Auger decay in the CS-CCSD wave function thus relies on 
the overlap between the valence orbitals which are emptied during the decay 
process and the virtual orbitals, in which the outgoing electron is quenched. 
The quantity $O$ from Eq. \eqref{eq:orb1} is presumably useful as well for 
analyzing the description of other states that decay by a two-electron process, 
i.\,e., Feshbach resonances in general.  

Our calculations show that the largest contribution to Im($O$) stems from 
$[\text{Im}(\varphi_{2\text{p},x}) \cdot 
\text{Re}(\varphi_{2\text{p},y}) + \text{Re}(\varphi_{2\text{p},x}) 
\cdot \text{Im}(\varphi_{2\text{p},y})] \cdot 
\text{Re}(\varphi_{n\text{d}, xy})$.
For the values in Tab. \ref{tab:orbs1}, this 
term is more than 10 times larger than those involving
$\text{Im}(\varphi_{n\text{d}, xy})$, 
in line with the nodal structures of the orbitals shown in Fig.
\ref{fig:orbitals}.

As a further step, we analyzed the molecular orbital coefficients of the 
d orbitals that deliver the largest contributions to the $^1$D decay width 
(5d, 6d, 7d). The results are compiled in Tab. \ref{tab:orbs2}; it is evident 
that basis functions with intermediate exponents between 1 and 10 are 
responsible for the largest share of the decay width.

This explains again the basis-set dependence documented in Tab. \ref{tab:cs}: 
Diffuse basis functions produce low-lying virtual orbitals that do not overlap 
with the occupied valence orbitals, while steep basis functions lead to virtual 
orbitals with high energy, whose overlap with the occupied orbitals cancels out 
due to oscillations. Only the intermediate virtual orbitals with an energy in 
the range of that of the emitted Auger electron overlap substantially with the 
valence orbitals and thus contribute to the Auger decay width.

\begin{table}
\caption{Real parts of molecular orbital coefficients of selected d$^\upalpha$ 
orbitals obtained from a CS-UHF calculation for Ne$^+$ (1$\text{s}_\upalpha^{-1}$)
using the aug-cc-pCV5Z basis set.$^\text{a}$}
\begin{ruledtabular}\begin{tabular}{lrrr}
Exponent & \multicolumn{3}{c}{Contribution to MO} \\
of bf & 5d &6d & 7d \\ \hline
212 & $-0.0009$ & $0.003$ & $-0.01$ \\
75.8 & $0.005$ & $-0.01$ & $0.06$ \\
27.0 & $-0.02$ & $0.07$ & $-0.2$ \\
9.84 & $0.1$ & $-0.2$ & \textbf{1.7} \\
3.84 & $-0.2$ & \textbf{1.7} & \textbf{--1.7} \\
1.50 & \textbf{1.6} & \textbf{--1.7} & $1.0$ \\
0.587 & \textbf{--1.4} & $0.8$ & $-0.4$ \\
0.213 & $0.5$ & $-0.3$ & $0.1$ \\
\end{tabular} \end{ruledtabular} \label{tab:orbs2} 
\footnotetext{The corresponding imaginary parts, which are much smaller, 
as well as the coefficients for the remaining d orbitals are available 
from the SI.}
\end{table}

%%%%%%%%%%%%%%%%%%%%%%%%%%%%%%%%%%%%%%%%%%%%%%%%%%%%%%%%%%%%%%%%%%%%%%%%%%%%

%$\text{Im}(\psi_{\text{rad, }2p\text{x}})$ and the most relevant
%$\text{Re}(\psi_{\text{rad, }n\text{d}_{xy}})$

%The imaginary part of this three-particle overlap thus has the form
%\begin{widetext}
%\begin{align}\nonumber
%    \text{Im}(O) = \int_0^\infty (&(\text{Re}(\psi_{\text{rad, 2p}_x}^{\upalpha
%*}(r))\cdot\text{Re}(\psi_{\text{rad, 2p}_y}^{\upbeta
%*}(r))-\text{Im}(\psi_{\text{rad, 2p}_x}^{\upalpha
%*}(r))\cdot\text{Im}(\psi_{\text{rad, 2p}_y}^{\upbeta *}(r)))
%\text{Im}(\psi_{\text{rad, }n\text{d}_{xy}}^\upalpha(r))\\
%    &-(\text{Im}(\psi_{\text{rad, 2p}_x}^{\upalpha
%*}(r))\cdot\text{Re}(\psi_{\text{rad, 2p}_y}^{\upbeta
%*}(r))+\text{Re}(\psi_{\text{rad, 2p}_x}^{\upalpha
%*}(r))\cdot\text{Im}(\psi_{\text{rad, 2p}_y}^{\upbeta
%*}(r)))\text{Re}(\psi_{\text{rad, }n\text{d}_{xy}}^\upalpha(r)))\cdot 4\uppi
%r^2\text{d}r.
%\end{align}
%\end{widetext}

%%%%%%%%%%%%%%%%%%%%%%%%%%%%%%%%%%%%%%%%%%%%%%%%%%%%%%%%%%%%%%%%%%%%%%%%%%%

\subsection{Total and partial Auger decay widths from complex basis functions} 
\label{sec:necbf}
Having identified the most important basis functions for the description of 
Auger decay in in CS-CC calculations, we are now in a position to conduct 
CBF-CCSD and CBF-EOMIP-CCSD calculations; the results are shown in Tab. 
\ref{tab:cbf}. We started by scaling all functions in the aug-cc-pCV5Z basis 
set and then proceeded by scaling fewer and fewer functions as illustrated 
in Tab. \ref{tab:cbf}. For technical reasons, the unscaled STO-2G basis had 
to be added to some calculations. Some additional results computed with other 
choices of scaled basis functions are available from the SI.

\begin{table}
\caption{Energies and half-widths of Ne$^+$ (1$\text{s}^{-1}$) computed 
with CBF-EOMIP-CCSD and CBF-$\Delta$CCSD by scaling different parts of the 
basis set. Energies in eV. Half-widths in meV.}
\begin{ruledtabular}\begin{tabular}{llcrr}
Method & Scaled bfs$^\text{a}$ & $\theta_\text{opt} / ^\circ$ & $\text{Re}(E)$ &
$\text{Im}(E)$ \\ \hline
\multicolumn{5}{c}{Basis set: STO-2G + aug-cc-pCV5Z} \\ \hline
$\Delta$CCSD & aug-cc-pCV5Z & 10 & 869.53 & $-109$ \\
EOMIP-CCSD & aug-cc-pCV5Z & 11 & 871.20 & $-116$ \\ 
EOMIP-CCSD & 4D, 2D, 0.6D, 0.2D, & 13 & 871.18 & $-118$ \\ 
 & 1P, 2S, 3F, 1F & \\ \hline
\multicolumn{5}{c}{Basis set: aug-cc-pCV5Z} \\ \hline
$\Delta$CCSD & 4D & 17 & 869.49 & $-100$ \\
EOMIP-CCSD & 10D, 4D, 2D & 18 & 871.21 & $-71$ \\
EOMIP-CCSD & 4D & 21 & 871.18 & $-89$ \\
EOMIP-CCSD & 4S & 24 & 871.04 & $-20$ \\
EOMIP-CCSD & 4D, 2D, 0.6D, 0.2D, & 6 & 871.08 & $-124$ \\ 
 & 1P, 2S, 3F, 1F \\ \hline
\multicolumn{5}{c}{Reference values} \\ \hline
\multicolumn{2}{l}{CS-$\Delta$CCSD / aug-cc-pCV5Z} & 12 & 869.53 & $-104$ \\
\multicolumn{2}{l}{CS-EOMIP-CCSD / aug-cc-pCV5Z} & 13 & 871.20 & $-114$ \\
\multicolumn{3}{l}{Fano} & 870.12~\cite{coreno99} & 
$-109$~\cite{skomorowski21b} \\
\multicolumn{3}{l}{Experiment} & 870.17~\cite{coreno99} & 
$-129$~\cite{muller17} \\
\end{tabular} \end{ruledtabular} \label{tab:cbf} 
\footnotetext{The numbers in the basis-set specification refer to the rounded 
exponent of the basis function in atomic units while the letter indicates the 
angular momentum.}
% compared to the values obtained with the CS method and experimental and
%theoretical reference values. 
\end{table}

The results in Tab. \ref{tab:cbf} demonstrate that CBF-$\Delta$CCSD and
CBF-EOMIP-CCSD both reproduce the CS reference values from Tab. \ref{tab:cs}
to an excellent degree when all basis functions are scaled. When we instead
scale only the three d shells with the largest contribution to $\Gamma/2$ 
(see Tab. \ref{tab:orbs2}), we still reproduce 95~\% of the $^1$D partial decay 
width from Tab. \ref{tab:cspw}. However, our results indicate that the 
scaled part of the basis set can be reduced further: With a single scaled
d shell (exponent = 3.844), we obtain 120~\% of the CS-CCSD value for the 
$^1$D partial width. The same procedure is also successful for the two $^1$S 
decay channels: Scaling one s function (exponent = 4.327) results in 120~\% 
of the CS-CCSD partial widths for these channels. Furthermore, by scaling 
a total of 8 s, p, d, and f shells, we recover the total decay width up to 
a few percent. 

It thus appears that one can compute partial decay widths with CBF methods by 
scaling only basis functions of the respective angular momentum. However, for 
decay into the $^1$P and $^3$P channels, we encountered convergence problems or 
obtained qualitatively incorrect results in calculations with a single scaled 
p shell. This problem presumably arises because decay into the $^1$P and $^3$P 
channels involves the occupied 2p orbital and additionally some virtual 
p orbitals. Complex scaling a p shell thus affects the description of the 
occupied 2p orbital, whereas for the $^1$D (2p$^{-2}$) and $^1$S 
(2p$^{-2}$) channels the involved occupied orbitals (2p) are of a different 
angular momentum than the relevant virtual orbitals (d or s, respectively). 
As documented in the SI, a solution to this problem is to add complex-scaled 
functions to a basis set instead of scaling functions that are already contained 
in the basis set. In this way, the consistency of the predefined basis 
is preserved. 
 
A further detail in Tab. \ref{tab:cbf} worth mentioning are the differences in 
$\theta_\text{opt}$ between the decay channels. Using Eqs. \eqref{eq:decomp1b} 
and \eqref{eq:decomp1}, the CS-CCSD partial widths are evaluated at the same 
$\theta$ and thus not at the respective $\theta_\text{opt}$ but this does not
seem to affect the quality of the partial widths obtained with this approach
as Tab.~\ref{tab:cspw} illustrates.

\subsection{Truncation of the complex-scaled basis set} \label{sec:trunc}
At this point, two questions still need to be answered to compute Auger 
decay rates in a black-box fashion with CBF methods: First, how to choose 
the complex-scaled functions that are added to a predefined basis set and, 
second, how to truncate the aug-cc-pCV5Z basis set that we used so far 
without compromising accuracy. 

\begin{figure} \centering
\includegraphics[width=0.93\linewidth]{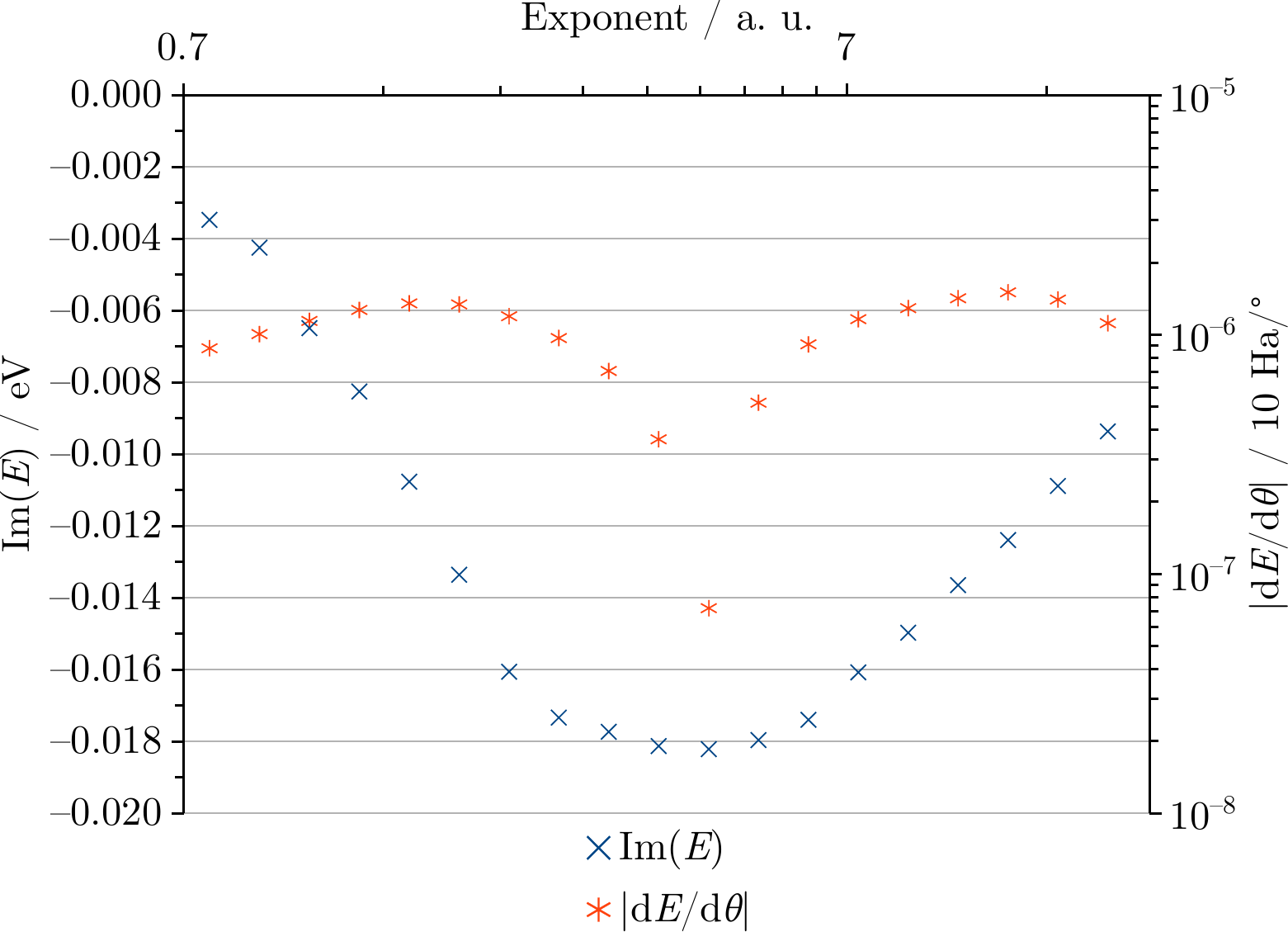} \\[0.3cm]
\includegraphics[width=0.93\linewidth]{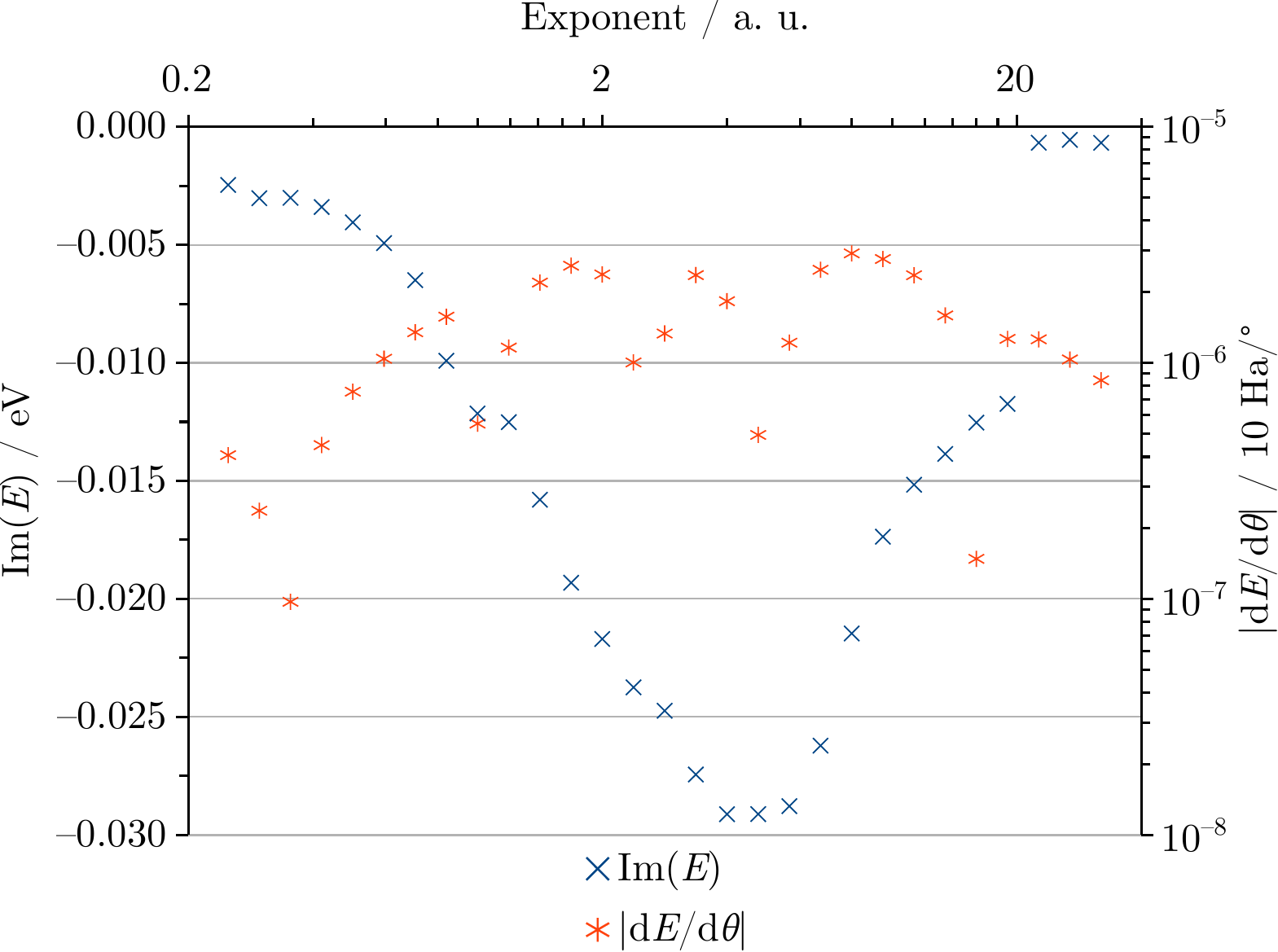} \\[0.3cm]
\includegraphics[width=0.93\linewidth]{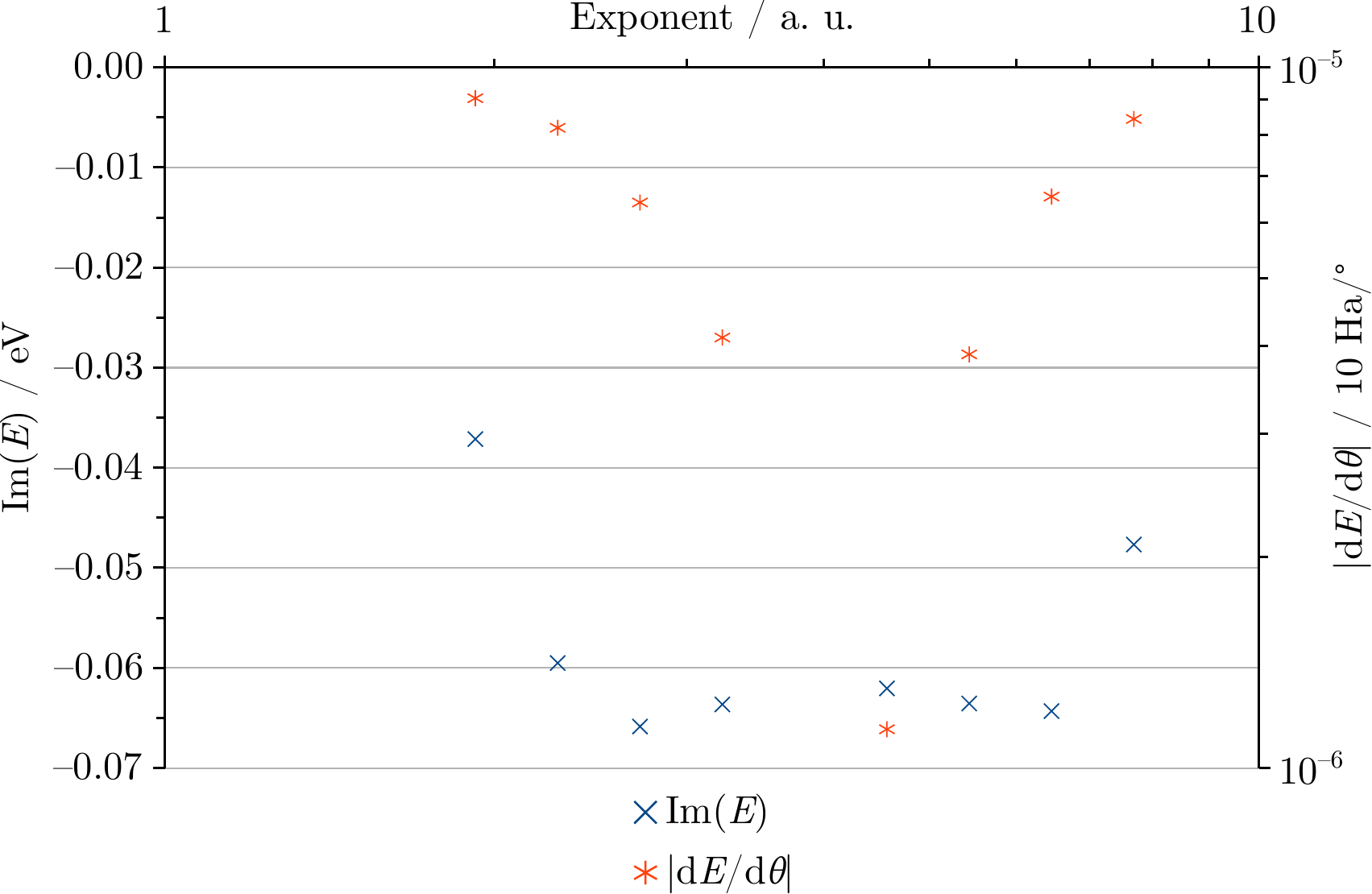}
\caption{Decay half-widths $-\text{Im}(E)$ and stabilizations
$|\text{d}E/\text{d}\theta|$ of Ne$^+$ (1s$^{-1}$) computed with
CBF-EOMIP-CCSD/aug-cc-pCV5Z as a function of the exponent of the single
complex-scaled shell that was added to the basis set. $\theta$ was optimized for
each exponent. The complex-scaled shell is an s, p, d shell in the upper, middle,
and lower panel, respectively.}
\label{fig:screen} \end{figure}

To investigate the requirements on the complex-scaled exponents, we performed 
series of CBF-EOMIP-CCSD calculations where we added one additional s, p, or 
d shell with varying exponent to the aug-cc-pCV5Z basis set. For each value 
of the exponent, a trajectory $E(\theta)$ was computed. This procedure is 
equivalent to optimizing the exponent of the extra shell in the complex 
number plane.\cite{moiseyev11} The resulting decay half-widths are presented 
in Fig. \ref{fig:screen} together with the value of $|\text{d}E$/d$\theta|$ 
at the respective $\theta_\text{opt}$. 

All three panels of Fig. \ref{fig:screen} have in common that the decay 
width is only captured when the exponent of the complex-scaled shell falls 
in a specific range spanning ca. one order of magnitude. Outside these ranges,
Im($E$) quickly approaches zero. The extrema in Im($E$) observed in Fig. 
\ref{fig:screen} at 18 meV (scaled s function, exponent 4.33), at 29 meV 
(scaled p shell, exponent 4.00), and at 66 meV (scaled d shell, exponent 
2.72) are very well in line with the partial widths obtained through 
decomposition of the CS-CCSD energy and also accurate estimates of the 
experimental values (see Tab. \ref{tab:cspw}). In the SI, we demonstrate 
through decomposition of the CBF-CCSD energy according to Eq.~\eqref{eq:decomp1}
that decay widths obtained as in Fig. \ref{fig:screen} indeed stem from only 
those channels that correspond to the angular momentum of the complex-scaled 
shell.

\begin{table} %Unrelated info: cc-pCV5Z = 145 bf, cc-pCVTZ (5sp) = 59 bf, cc-pCVDZ (5sp) = 42 bf
\caption{Partial decay half widths of Ne$^+$ (1$\text{s}^{-1}$) computed with 
CBF-EOMIP-CCSD and various basis sets. All values in meV.}
\begin{ruledtabular}\begin{tabular}{lclrrrr}
 & \# of & Complex & \multicolumn{4}{c}{$-\,\text{Im}(E)$} \\
Basis set & b.f.s & shells$^\text{a}$ & D & P &
S & all$^b$ \\ \hline
%aug-cc-pCV5Z & 181 & 1 optimized & 63 & 13 & 18 & 93 \\
aug-cc-pCV5Z & 181 & opt. & 64 & 31 & 19 & 114 \\
aug-cc-pCVQZ & 109 & opt. & 67 & 51 & 15 & 132 \\
aug-cc-pCVQZ & 109 & from 5Z & 67 & 19 & 20 & 106 \\
aug-cc-pCVTZ & 59  & opt. & 90 & 4 & 36 & 130 \\
aug-cc-pCVTZ & 59  & from 5Z & 77 & 3 & 13 & 93 \\
aug-cc-pCVTZ (unc.)$^c$ & 71 & opt. & 89 & 101 & 18 & 208 \\
aug-cc-pCVTZ (unc.)$^c$ & 71 & from 5Z & 80 & 42 & 18 & 140 \\
6-311+G(3df) (unc.)$^c$ & 52 & opt. & 118 & 59 & 60 & 237 \\
6-311+G(3df) (unc.)$^c$ & 52 & from 5Z & 82 & 20 & 29 & 131 \\ 
%6-311+G(3df) (unc.)$^c$ & 52 & from 5Z$^e$ & 78 & 17 & 37 & 131 \\
aug-cc-pCVTZ (5sp)$^d$ & 75 & opt. & 91 & 31 & 18 & 139 \\
aug-cc-pCVTZ (5sp)$^d$ & 75 & from 5Z & 80 & 34 & 18 & 131 \\ \hline
%aug-cc-pCVTZ (5sp)$^d$ & 75 & from 5Z$^e$ & 76 & 34 & 18 & 127 \\
\multicolumn{3}{l}{CS-$\Delta$CCSD / aug-cc-pCV5Z, from Tab. \ref{tab:cspw}} & 
75 & 35 & 17 & 126 \\
\multicolumn{3}{l}{Fano / EOM-CCSD, from Ref. \citenum{skomorowski21b}} & 
59 & 32 & 19 & 109 \\
\multicolumn{3}{l}{Experiment, from Refs. \citenum{muller17,albiez90}} & 
79 & 30 & 20 & 129 \\
\end{tabular} \end{ruledtabular} \label{tab:scr} 
\footnotetext{Two complex-scaled shells are used in all calculations, 
their exponents are either optimized for the aug-cc-pCV5Z basis set or the
basis set given in the first column of the table. The values are available
from the SI. See text for further explanation.} 
\footnotetext{Evaluated as sum of the values in the 3 preceding columns. 
See Sec. \ref{sec:nepw} for further discussion.}
\footnotetext{Fully uncontracted basis set.}
\footnotetext{aug-cc-pCVTZ with all s and p shells replaced by those from
aug-cc-pCV5Z.}
\end{table}

% Two complex-scaled shells are sufficient 
To determine to what degree the unscaled basis can be truncated, we performed 
CBF-EOMIP-CCSD calculations with different Dunning and Pople basis sets 
supplemented by two complex scaled shells. The results are summarized in Tab.
\ref{tab:scr} and in the SI. The exponents of the extra shells are chosen such
that they minimize $|\text{d}E$/d$\theta|$ in the aug-cc-pCV5Z basis or,
alternatively, in the basis set used in the respective calculation; all exponents
are available in the SI. 

Tab. \ref{tab:scr} illustrates that the results quickly deteriorate when the 
basis set is truncated in a straightforward way. The total decay width is 
somewhat insensitive, but the branching ratios are qualitatively wrong already 
with aug-cc-pCVQZ independent of reoptimizing the exponents of the two 
complex-scaled shells. Somewhat better results are obtained with the uncontracted 
aug-cc-pCVTZ basis set, but uncontraction does not seem to give reliable results 
universally: With the uncontracted 6-311+G(3df) basis set, which was recently 
identified as a good compromise between accuracy and computational cost for 
core-ionization energies,\cite{sarangi20} the total decay width is acceptable
but the branching ratios are still distorted.  

However, there is another way to truncate the basis set: aug-cc-pCV5Z 
includes a lot of shells with $L>2$ that neither contribute to the occupied 
orbitals nor are they involved in Auger decay. Since we thus need primarily 
s- and p shells to improve the description, we opted to replace the s- and p-
shells of aug-cc-pCVTZ with those from aug-cc-pCV5Z. The resulting basis is 
denoted ``aug-cc-pCVTZ (5sp)'' and capable of producing accurate branching 
ratios and total widths as Tab. \ref{tab:scr} demonstrates. 

Notably, reoptimization of the exponents of the complex-scaled shells does 
not help but impairs the results. This shows again that the exponents of these
extra functions bear a meaning independent of the unscaled basis set; 
specifically, they are related to the energy of the Auger electron. This 
conclusion is analogous to the analysis of the CS-CCSD wave function in 
Tab. \ref{tab:orbs1}.

\section{Molecular Auger Decay} \label{sec:molres}
\subsection{General considerations} \label{sec:genmol}

Several aspects deserve attention when applying CBF-CC methods to Auger decay 
in molecules instead of atoms: Since for molecules $[\hat{H}, \hat{L}^2] 
\neq 0$, partial widths cannot be determined through scaling only shells of 
a particular angular momentum. \textit{A priori}, it is unclear if this mixing 
of different angular momenta raises or lowers the requirements towards the 
basis set. In addition, we need a procedure to determine the exponents of 
the complex-scaled shells for different atoms without the costly optimization 
that we carried out for Ne in Fig. \ref{fig:screen}. A related question is 
whether it is sufficient to add complex-scaled shells only to the basis set 
of the atom with the core vacancy or if these extra shells are needed at the 
other atoms as well. 

In the following, we investigate these aspects by the examples of core-ionized 
H$_2$O, N$_2$, and C$_6$H$_6$. We find that choosing the exponents of the 
complex-scaled shells is most critical. As established in Sec. \ref{sec:atres}, 
the virtual orbitals constructed from these basis functions need to overlap 
with the occupied orbitals that are emptied during the decay process in order 
to describe the quenching of the outgoing electron. Since the diffuseness of 
the virtual orbitals strongly depends on the nuclear charge, the exponents 
of the complex-scaled shells need to be chosen carefully. 

In this work, we use the geometric mean $\bar{\zeta}=(\prod_i^N \zeta_i)^{1/N}$ 
of the exponents in a basis set as a measure of diffuseness. For basis functions 
contracted from $M$ primitive Gaussians with exponents $\eta_j$, we assign 
to them a $\zeta_i$ calculated as $\zeta_i=(\prod_j^M \eta_j)^{1/M}$. We then 
use the ratio between the $\bar{\zeta}$ values of different atoms as a scaling 
factor to adjust the exponents of the complex-scaled shells starting from the
values for Ne established in Sec. \ref{sec:trunc}. For the aug-cc-pCV5Z basis,
the values of $\bar{\zeta}$ are 0.64, 2.55, 3.52, 4.48, and 6.99 for H, C, N, O,
and Ne, respectively, but differences between basis sets are small. All exponents
used in our calculations on H$_2$O, N$_2$, and C$_6$H$_6$ as well as the molecular
structures can be found in the SI.

\subsection{Water} \label{sec:h2o}
As a first example, we consider the core-ionized state of water, which has 
the electronic configuration 1a$_1^1$2a$_1^2$1b$_2^2$3a$_1^2$1b$_1^2$. There 
are 16 main decay channels that can be distinguished by the spin and spatial 
symmetry of the dicationic target state. We performed CBF-$\Delta$CCSD 
and CBF-EOMIP-CCSD calculations with the basis sets aug-cc-pCVTZ (5sp) 
and 6-311+G(3df) (unc.) introduced in Sec. \ref{sec:trunc}. Both basis 
sets were augmented by 2 complex-scaled s, p, and d shells.% whose exponents 
%were determined using the procedure from Sec. \ref{sec:genmol} starting 
%from the exponents for Ne in the aug-cc-pCV5Z (5sp) basis set. 

\begin{table}
\caption{Energies and half-widths of H$_2$O$^+$ (1s$^{-1}$) computed with 
CBF-EOMIP-CCSD and CBF-$\Delta$CCSD. Energies in eV. Half-widths
in meV.}
\begin{ruledtabular}\begin{tabular}{llcrr}
Unscaled & Complex- & $\theta_\text{opt} / ^\circ$ & $\text{Re}(E)$ &
$\text{Im}(E)$ \\ 
basis set & scaled shells & \multicolumn{3}{c}{EOMIP-CCSD} \\ \hline
6-311+G(3df) (unc.) & 2$\times$(spd) on O, H & 13 & 541.4 & $-$85 \\
6-311+G(3df) (unc.) & 2$\times$(spd) on O & 14 & 541.3 & $-$82 \\
aug-cc-pCVTZ (5sp) & 2$\times$(spd) on O, H & 26 & 541.4 & $-$75 \\
aug-cc-pCVTZ (5sp) & 2$\times$(spd) on O & 29 & 541.4 & $-$78 \\ \hline
 &  & \multicolumn{3}{c}{$\Delta$CCSD} \\ \hline
6-311+G(3df) (unc.) & 2$\times$(spd) on O, H & 14 & 539.6 & $-$90 \\
aug-cc-pCVTZ (5sp) & 2$\times$(spd) on O, H & 41 & 539.7 & $-$78 \\ \hline
 &  & \multicolumn{3}{c}{Reference values} \\ \hline
\multicolumn{3}{l}{EOM-CCSD + Fano \cite{skomorowski21b}} & 
--- & $-$61 \\
\multicolumn{3}{l}{MRCI + Fano \cite{sankari03,inhester12,inhester14}} & 
539.6 & $-$73 \\
\multicolumn{3}{l}{Experiment \cite{sankari03}} & 539.8 & $-$80(2) \\
\end{tabular} \end{ruledtabular}
\label{tab:h2o} 
\end{table}
% basis sets, compared to experimental and theoretical reference values.

The results can be found in Tab.~\ref{tab:h2o}. For energies, EOMIP-CCSD
yields values that are ca. 2 eV higher as compared to $\Delta$CCSD, which is
similar to neon (see Tab. \ref{tab:cs}) and has been discussed elsewhere in 
detail.\cite{zheng19,matthews20} The decay widths obtained with the 
two basis sets differ by ca. 10~\% but the stabilization is much 
better with aug-cc-pCVTZ (5sp) ($|\text{d}E$/d$\theta|$ = $7.5 \cdot 
10^{-6} /^\circ$) than with 6-311+G(3df) (unc.) ($|\text{d}E$/d$\theta|$ 
= $1.3 \cdot 10^{-4}/^\circ$). Removal of the complex-scaled shells 
at the hydrogen atoms makes only a marginal impact in both bases.

To compute the partial widths of the 16 decay channels, we applied Eq. 
\eqref{eq:decomp1} to the CBF-CCSD wave function of H$_2$O$^+$ (1s$^{-1}$). 
Tab. \ref{tab:h2opw} shows results obtained with the 
aug-cc-pCVTZ (5sp) basis set, which delivers accurate branching ratios 
for the neon atom  (see Table~\ref{tab:scr}), and with the 6-311+G(3df) 
(unc.) basis set used in Ref. \citenum{skomorowski21b}. In both cases, 
CBFs were placed on all atoms. Our results obtained in the two different 
basis sets agree well with each other and also with those from the 
Fano-EOM-CCSD and Fano-MRCI treatments from Refs. \citenum{skomorowski21b} 
and \citenum{inhester12}, respectively. Interestingly, 
the results obtained with the 6-311+G(3df) (unc.) basis set are very 
close to those calculated with the Fano-EOM-CCSD method within the same
basis set\cite{skomorowski21b} while those calculated with aug-cc-pCVTZ 
(5sp) are closer to the Fano-MRCI results \cite{inhester12,inhester14} 
where the cc-pVTZ basis set was used.

%to establish the best possible comparison to the
%available results from the literature (Ref. \citenum{skomorowski21b}) where this
%basis set was used as well.
%can be identified as a methodic error stemming from the EOM-CCSD parametrization. 
%This state was reported before~\cite{agren80} to be highly mixed with multiple 
%shake-up states, reducing its intensity, but this mixing is not included in the 
%EOM-CCSD wave function and can be described much better by a MRCI treatment.

All approaches find that decay into singlet states is much more probable 
than into triplet states, the latter account for a mere 5-10~\% of the 
total decay width. There is also good agreement about individual 
channels with one conspicuous exception: Our computations and Ref. 
\citenum{skomorowski21b} assign to the 2a$_1$2a$_1$ channel the 
largest partial width, whereas this channel is of minor importance 
in the Fano-MRCI treatment. This may be due to the 
mixing of this state with multiple shake-up states,\cite{agren80} 
which can be described better by MRCI than by EOM-CCSD or $\Delta$CCSD.

\begin{table}
\caption{Partial decay widths of H$_2$O$^+$ (1s$^{-1}$) computed 
with different methods. All values in meV.}
\begin{ruledtabular}\begin{tabular}{lrrrr}
Decay & CBF- & CBF- & Fano &  Fano \\ 
channel & $\Delta$CCSD$^\text{a}$ & $\Delta$CCSD$^\text{b}$ & MRCI$^\text{c}$ &
EOM-CCSD$^\text{d}$ \\ \hline
$3\text{a}_1 1\text{b}_1$ (triplet) & 0.3 & 0.2 & 0.4 & 0.5 \\
$1\text{b}_1 1\text{b}_1$ & 12.2 & 18.0 & 19.0 & 13.3 \\
$3\text{a}_1 1\text{b}_1$ (singlet) & 14.8 & 19.6 & 18.0 & 12.7 \\
$1\text{b}_1 1\text{b}_2$ (triplet) & 0 & 0 & 0 & 0 \\
$3\text{a}_1 3\text{a}_1$ & 10.0 & 12.2 & 13.1 & 8.9 \\
$1\text{b}_1 1\text{b}_2$ (singlet) & 12.1 & 15.7 & 15.2 & 10.7 \\
$3\text{a}_1 1\text{b}_2$ (triplet) & 0.2 & 0.2 & 0.3 & 0.4 \\
$3\text{a}_1 1\text{b}_2$ (singlet) & 10.1 & 13.4 & 13.2 & 9.5 \\
$1\text{b}_2 1\text{b}_2$ & 7.0 & 8.7 & 9.8 & 7.1 \\
$2\text{a}_1 1\text{b}_1$ (triplet) & 3.6 & 2.8 & 3.0 & 4.1 \\
$2\text{a}_1 3\text{a}_1$ (triplet) & 3.3 & 2.5 & 2.6 & 3.8 \\
$2\text{a}_1 1\text{b}_2$ (triplet) & 2.6 & 2.2 & 1.6 & 2.9 \\
$2\text{a}_1 1\text{b}_1$ (singlet) & 13.9 & 9.6 & 10.0 & 9.5 \\
$2\text{a}_1 3\text{a}_1$ (singlet) & 15.1 & 12.7 & 11.0 & 13.6 \\
$2\text{a}_1 1\text{b}_2$ (singlet) & 2.6 & 6.8 & 6.6 & 6.3 \\
$2\text{a}_1 2\text{a}_1$ & 19.0 & 21.6 & 4.1 & 15.3 \\ \hline 
All & 129.9 & 142.5 & 145.6 & 121.7 \\ 
\end{tabular} \end{ruledtabular} \label{tab:h2opw}
\footnotetext{This work. Computed using the aug-cc-pCVTZ (5sp) 
basis set and Eq.~\eqref{eq:decomp1}.}
\footnotetext{This work. Computed using the 6-311+G(3df) (unc.) basis set 
and Eq.~\eqref{eq:decomp1}.}
\footnotetext{From Ref. \citenum{inhester12,inhester14}.}
\footnotetext{From Ref. \citenum{skomorowski21b}.}
\end{table}

It should be noted that the CBF-$\Delta$CCSD partial widths in Tab. 
\ref{tab:h2opw} add up to a value of 146.2 meV, i.\,e., not the value 
reported there as ``all'' (142.5 meV), which was 
obtained by removing all Auger-like transitions at once from $T$ 
and $\Lambda$ before evaluating Eq. \eqref{eq:decomp1}. Moreover, 
this value is different from the total width of H$_2$O$^+$ (1s$^{-1}$) 
in Tab. \ref{tab:h2o} (180~meV). The reasons are
the same that we discussed in detail for neon in Sec. \ref{sec:nepw}. 
We point out that the second discrepancy is not entirely spurious; 
the total width from Tab.~\ref{tab:h2o} contains contributions related 
to satellite states and interchannel coupling (see Sec. \ref{sec:nepw}). 
Remarkably, the partial widths from Refs. \citenum{skomorowski21b} and 
\citenum{inhester12} computed using Fano's theory also do not add up 
to the respective total widths, presumably because of contributions 
from satellite states as well.

\begin{figure} \centering
\includegraphics[width=0.95\linewidth]{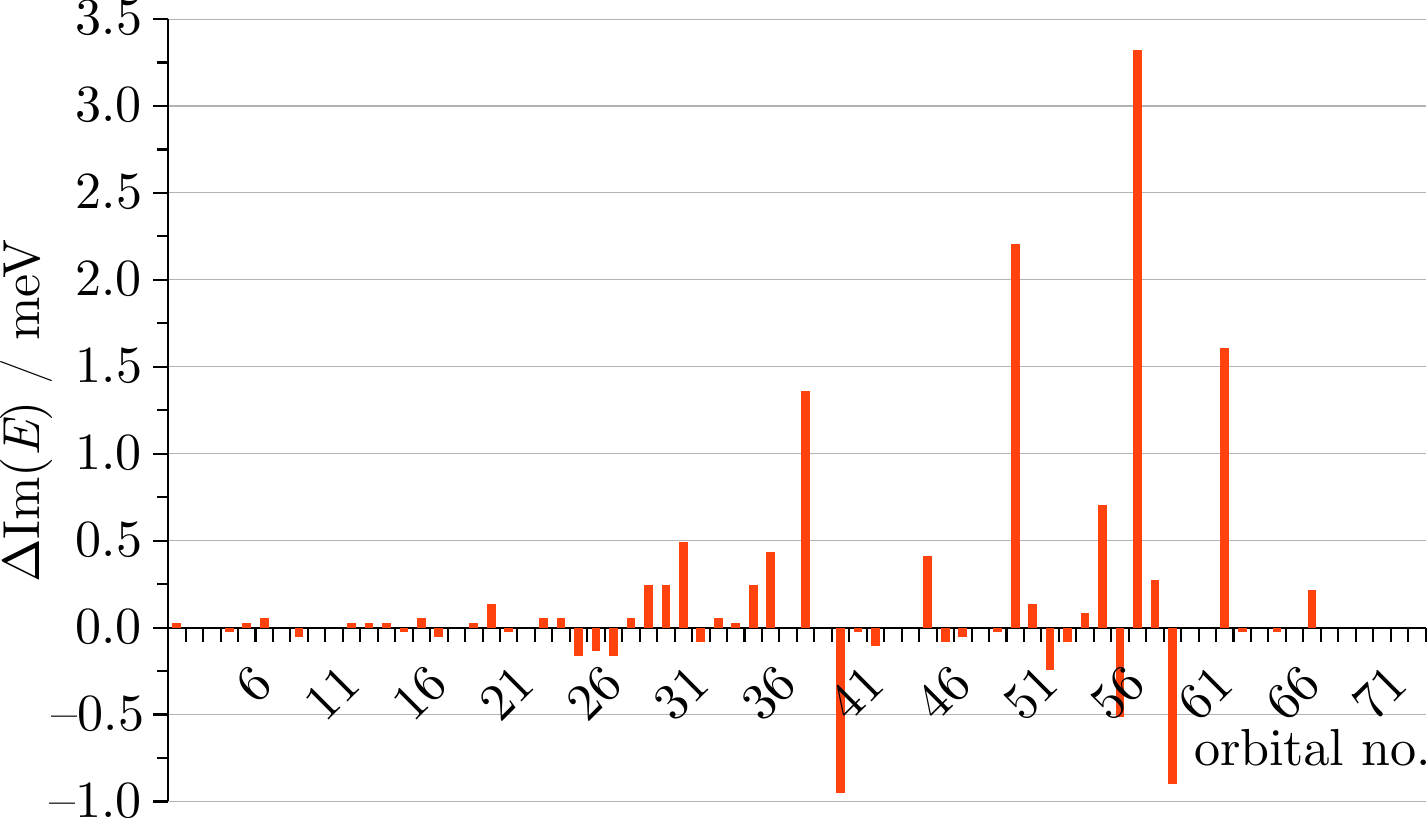}
\caption{Contributions of virtual orbitals of a$_1$ symmetry to the 
CBF-CCSD decay width of the $2\text{a}_1 2\text{a}_1$ channel of 
H$_2$O$^+$ (1s$^{-1}$). The analysis was done with the aug-cc-pCVTZ (5sp) 
basis and 2 complex-scaled s, p, and d shells on all atoms.}
\label{fig:h2odec}
\end{figure}

As a further analysis step, we decomposed the decay width of the 
$2\text{a}_1 2\text{a}_1$ channel into contributions from individual 
virtual orbitals. The result is shown in Fig. \ref{fig:h2odec}. Similar 
to neon (see Tab. \ref{tab:orbs1}) a few orbitals account for the 
largest share of $\Gamma$. The most important orbital 59$a_1$, which 
contributes 3.3 meV, has an energy of 475 eV; a value that represents 
a good approximation to the energy of the emitted Auger electron 
(456~eV).\cite{moddeman71}

\subsection{Dinitrogen} \label{sec:n2}
As a second example, we examined the core-ionized states of the nitrogen 
molecule. The neutral ground state has the electronic configuration 
$1\upsigma_g^2 1\upsigma_u^2 2\upsigma_g^2 2 \upsigma_u^2 1\uppi_g^4 
3\upsigma_g^2$. Two core-ionized states $^2\Sigma_g^+$ and $^2\Sigma_u^+$ 
can be distinguished depending on whether an electron is removed from 
the 1$\upsigma_g$ or the 1$\upsigma_u$ orbital. The energy splitting 
between these two orbitals is very small (ca. 80~meV in an RHF calculation) 
so that one expects the two core-ionized states to overlap in terms 
of their widths. This aspect has been discussed elsewhere before.\cite{
lindle84,hergenhahn01,semenov05,sorensen08}

\begin{table}
\caption{Energies and half-widths of the $^2\Sigma_g^+$ and 
$^2\Sigma_u^+$ core-ionized states of N$_2^+$ computed with 
CBF-EOMIP-CCSD. Energies in eV. Half-widths in meV.}
\begin{ruledtabular}\begin{tabular}{llcrr}
Unscaled & Complex- &  &  &  \\
basis set & scaled shells$^\text{a}$ &
$\theta_\text{opt} / ^\circ$ & 
$\text{Re}(E)$ & $\text{Im}(E)$ \\ \hline
\multicolumn{5}{c}{$^2\Sigma_g^+$ state} \\ \hline
aug-cc-pCV5Z & 3$\times$(spd) & 18 & 411.28 & $-$55 \\
aug-cc-pCVQZ & 3$\times$(spd) & 12 & 411.29 & $-$50 \\
aug-cc-pCVTZ & 3$\times$(spd) & 8 & 411.24 & $-$58 \\
aug-cc-pCVTZ (5sp) & 3$\times$(spd) & 22 & 411.24 & $-$63 \\ \hline
\multicolumn{3}{c}{Experiment} & 409.96$^\text{b}$ &
$-$58$^c$ \\ \hline
\multicolumn{5}{c}{$^2\Sigma_u^+$ state}\\ \hline
aug-cc-pCV5Z & 3$\times$(spd) & 25 & 411.19 & $-$55 \\
aug-cc-pCVQZ & 3$\times$(spd) & 12 & 411.19 & $-$53 \\
aug-cc-pCVTZ & 3$\times$(spd) & 8 & 411.14 & $-$54 \\
aug-cc-pCVTZ (5sp) & 3$\times$(spd) & 19 & 411.14 & $-$63 \\ \hline
\multicolumn{3}{c}{Experiment} & 409.88$^\text{b}$ & $-$62$^\text{c}$ \\
\end{tabular} \end{ruledtabular} \label{tab:n2} 
\footnotetext{Exponents determined according to Sec. \ref{sec:genmol}.}
\footnotetext{From Ref. \citenum{sorensen08}, vibrationally averaged.}
\footnotetext{From Ref. \citenum{semenov05}, vibrationally averaged.}
\end{table}

Our CBF-EOMIP-CCSD results in Tab. \ref{tab:n2} confirm that the two resonances 
overlap indeed. The widths of both states are computed to be ca. 120 meV in 
good agreement with experimental values,\cite{semenov05} whereas the energy 
gap is 100 meV. The experimental value for the energy gap is 80 meV\cite{
sorensen08} meaning the overlap is somewhat more pronounced. The ionization 
energies themselves are again systematically overestimated similar to what 
we found for neon and water.

We found that it is necessary to scale three s, p, and d 
shells in order to obtain converged results for the decay widths. When doing 
so, a standard basis set such as aug-cc-pCVTZ is already able to capture most 
of the total decay width. In contrast, when scaling only two shells, which 
works well for water, the decay widths deviate by more than 20\% from the 
experimental values and no basis set convergence is observed when going 
from aug-cc-pCVTZ to aug-cc-pCV5Z. This is documented in the Supporting 
Information. One may speculate that the need to scale a third shell is 
related to the presence of two heavy nuclei in N$_2$. The target states 
of Auger decay are presumably not well described when only two shells are 
complex scaled.

A further aspect of the core-ionized states of N$_2$ is vibrational 
progression resulting from the dependence of the energy and decay width 
on the bond length.\cite{lindle84,hergenhahn01,semenov05,sorensen08} 
The experimental values for $\Gamma/2$ in Tab. \ref{tab:n2} are vibrationally 
unresolved while our theoretical results in the same table are computed 
at $R$(NN)= 1.1 \AA\ and do not account for any vibrational effects. 

\begin{figure} 
\includegraphics[width=0.95\linewidth]{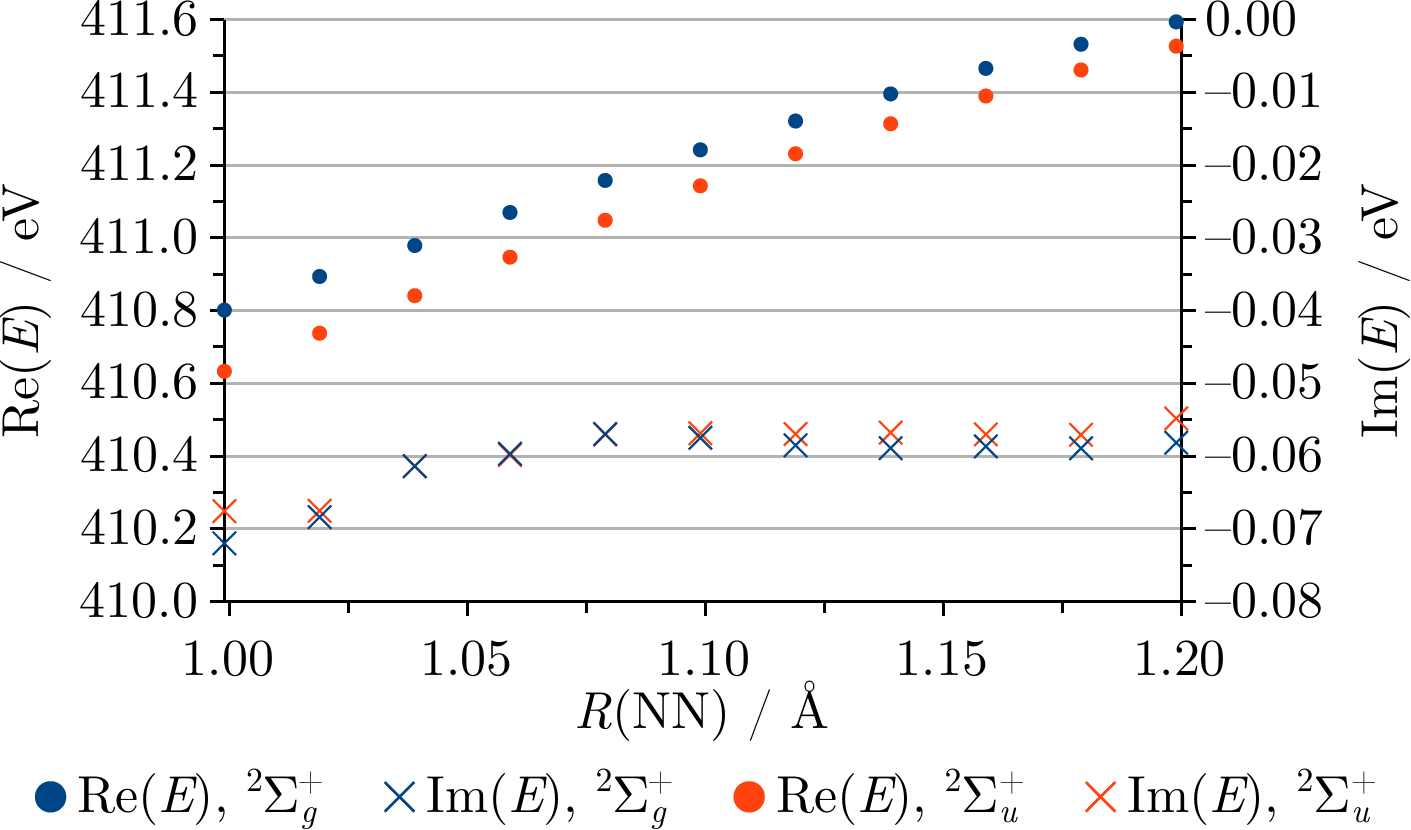}
\caption{Dependence of energies and decay widths of the $^2\Sigma_g^+$ 
and $^2\Sigma_u^+$ core-ionized states of N$_2^+$ on the NN bond distance 
computed with CBF-EOMIP-CCSD/aug-cc-pCVTZ (5sp) and three
complex-scaled s, p, and d shells on each atom.}\label{fig:n2} 
\end{figure}

To get an estimate of the dependence of $E$ and $\Gamma$ on the bond length, 
we recomputed these quantities in the range $R$(NN) = 1.00--1.20 \AA. The 
results in Fig. \ref{fig:n2} illustrate that the widths of 
both resonances depend only weakly on $R$(NN), only a slight increase is 
observed at shorter bond lengths. This finding is consistent with the core 
orbitals not participating in the bond between the two nitrogen atoms and 
can be contrasted with valence resonances such as the $^2\Pi_g$ state of 
N$_2^-$ where $\Gamma$ depends strongly on the molecular structure.\cite{
jagau17} In contrast to the decay width, the ionization energies change 
by ca. 1 eV in the range of $R$(NN) that we investigated. Also, the energy 
gap between the two states decreases at stretched bond lengths. 

\subsection{Benzene} \label{sec:c6h6}
To demonstrate the applicability of our approach to larger molecules, we 
investigated the lowest core-ionized state of benzene using CBF-EOMIP-CCSD 
and the 6-311+G(3df) (unc.), aug-cc-pCVTZ, and aug-cc-pCVTZ 
(5sp) basis sets. We placed 2 or 3 complex-scaled s, p, 
and d shells at all carbon atoms. The resulting numbers 
of basis functions and compute times as well as results of all computations 
can be found in Tab. \ref{tab:c6h6}.

\begin{table}
\caption{Energies and half-widths of the $^2\text{A}_{1g}$
core-ionized state of C$_6$H$_6^+$ computed with CBF-EOMIP-CCSD.}
\begin{ruledtabular}\begin{tabular}{lrrr}
Unscaled & 6-311+G(3df) & aug-cc- & aug-cc- \\ 
basis set & (unc.) & pCVTZ & pCVTZ (5sp) \\ \hline 
Complex- & 2$\times$(spd) & \multicolumn{2}{c}{3$\times$(spd)} \\
scaled shells & at all C atoms & \multicolumn{2}{c}{at all C atoms} \\ \hline
Basis functions & 450 & 654 & 798 \\
Compute time$^\text{a}$/h & 4.6 & 16.4 & 45.8 \\
$\theta_\text{opt}$ / $^\circ$ & 18 & 8 & 13 \\
$\text{d}E/\text{d}\theta$/Hartree/$^\circ$ & $3.4 \cdot 10^{-5}$ & $1.5 \cdot 10^{-4}$ & $2.0\cdot 10^{-6}$ \\
Re($E$)/eV & 291.88 & 291.98 & 291.99 \\
Im($E$)/meV & $-$58.5 & $-$37.8 & $-$41.6 \\
\end{tabular} \end{ruledtabular} \label{tab:c6h6} 
\footnotetext{Compute time for one complex energy on an Intel 
Xeon E5-2667 v4 CPU using 16 cores.}
%scaling 5.16
\end{table}

The results exhibit a similar trend as those for N$_2$:
aug-cc-pCVTZ and aug-cc-pCVTZ (5sp) yield almost identical ionization energies,
again somewhat higher than the experimental value of 290.42 eV\cite{rennie00}. 
The decay widths differ by about 10 \% from each other, which is also reminiscent 
of N$_2$, whereas the 6-311+G(3df) (unc.) basis produces a 50\% larger width. 
Given that aug-cc-pCVTZ (5sp) yields the smallest value for $\text{d}E/\text{d}\theta$ 
and also led to the best agreement with reference values for the other examples 
discussed in the previous sections, we conclude that our final result for the 
Auger decay width of benzene is 83 meV. This value is in good agreement with 
the decay width of core-ionized methane;\cite{skomorowski21b} a rigorous 
assessment is, however, difficult because no experimental or theoretical 
values are available for the decay width of benzene although the Auger 
spectrum has been studied theoretically before.\cite{tarantelli87}

We note that it takes almost two days on a state-of-the-art 
16-core machine to compute one complex EOMIP-CCSD energy in the aug-cc-pCVTZ 
(5sp) basis set (798 basis functions). Since 10 to 20 computations are 
necessary to determine the optimal scaling angle, this illustrates the 
size of the calculations that are possible with our current hardware.

\section{Conclusions and Outlook} \label{sec:conc} 
We have shown how to compute total and partial Auger decay widths in the 
framework of complex-variable coupled-cluster theory. We discussed the 
evaluation of these quantities based on CCSD and EOMIP-CCSD wave functions 
using complex scaling of the Hamiltonian or, alternatively, of parts of 
the basis set. The latter approach extends the formalism of complex scaling 
to molecular resonances and, in addition, is superior in terms of numerical 
performance. This manifests itself in smaller decay widths 
of bound states, whose lifetime is infinite in the complete basis-set limit,
and faster convergence of the HF and CCSD equations.

In complex-variable methods, the total decay width is obtained as imaginary 
part of the eigenvalue of a non-Hermitian Hamiltonian without the need 
to make any assumption about the wave function of the emitted electron. 
Our applications of complex-scaled basis functions to
core-ionized states of Ne, H$_2$O, and N$_2$ demonstrated excellent agreement
for total Auger decay widths %among our different approaches as well as 
with experimental and previous theoretical investigations
with errors of only a few percent. We also reported the
first value for the Auger decay width of core-ionized C$_6$H$_6$; the result
is in good agreement with the Auger decay width of CH$_4$. A caveat regarding
the accuracy of our results is that the basis-set requirements of 
complex-scaled calculations on core-vacant states are not yet fully 
explored. While two complex-scaled s, p, and d shells appear to be 
sufficient for describing Auger decay of Ne and H$_2$O, three sets 
are required for N$_2$.

We gained access to partial decay widths and branching ratios by decomposing the
imaginary part of the CCSD or EOM-CCSD energy, respectively, in terms of
individual amplitudes. This analysis illustrated that --as one would expect-- the
largest share of the decay width is delivered by those excitations, which are
removed from the wave function in CVS methods. However, other excitations yield
non-negligible contributions in EOMIP-CCSD, which complicates the analysis. In
addition, there is a nonadditivity of the partial widths in our approach. Overall,
we found that partial widths computed with $\Delta$CCSD are more reliable than
those computed with EOMIP-CCSD.

The analysis of the imaginary part of the energy also gave insight into the 
requirements towards the basis set that the treatment of Auger decay poses. 
We found that it is sufficient to add 1--3 complex-scaled s, p,
and d shells to an unscaled basis set that is suitable for the treatment of
core-vacant states. The exponents of these extra shells need to be chosen
carefully to capture the decaying character of a core-ionized state, but their
values can be estimated from the energy of the emitted Auger electron. This means
in effect that the complex-scaled shells have exponents in the range 1--10, which
is in contrast to resonances that decay by emission of slow electrons, where extra
diffuse shells are pivotal. 
%discussion of method to obtain exponents? similarity to real scaling?

% Auger decay of core-ionized states and related processes 
%such as resonant Auger decay and double Auger decay 
We consider our work a critical extension of CC theory for core-vacant states. 
In our view, the prospects for applying complex-variable methods to 
core-vacant states are rather bright. A particular strength 
of complex-variable methods is that they offer a unified treatment of different 
types of resonances and are equally applicable to Feshbach and shape resonances. 
The latter are relevant for low-energy electron attachment and tunnel ionization 
but also of interest in the context of X-ray spectroscopy.\cite{
sorensen08} Alternative approaches for decaying states based on Fano's theory work 
well for Feshbach resonances but face problems when applied to shape resonances. 
Their main advantage is lower computational cost; all relevant states are modeled 
as bound states and the decay is treated separately afterwards. 
One may speculate that methods based on Fano's theory will emerge 
as superior for cases where the partition of the Hilbert space into a bound and a 
continuum part poses no problems. The main advantage of complex-scaled calculations, 
in contrast, is that they can be used in a black-box fashion because no explicit 
treatment of the continuum is necessary. The need for the optimization of the scaling 
angle, however, increases the computational cost. In this work, we usually found 
10 calculations to be sufficient to determine it to sufficient accuracy. The largest 
calculation that we carried out comprised 798 basis functions.
%Together with the relatively modest basis-set 
%requirements of CBF-CC methods, this enables applications to medium-sized 
%molecules such as benzene. 

In order to treat larger systems with complex-scaled methods, 
two strategies appear worthwhile to pursue: On the one hand, our approach 
can be easily adapted to related electronic-structure methods such as the 
second-order CC model\cite{christiansen95} or algebraic diagrammatic construction 
(ADC) schemes\cite{schirmer82} which entail lower computational cost. On the 
other hand, complex-variable CCSD and EOMIP-CCSD can be combined with quantum 
embedding.\cite{parravicini21} 

We add that our approach to extract partial widths from complex-variable 
calculations is not specific to Auger decay. In complex systems, not only 
Auger decay is of interest but also related non-local phenomena such as 
intermolecular Coulombic decay and electron-transfer mediated decay. We 
anticipate that our approach can be applied to these processes as well. 
Finally, we mention that nuclear motion will need to be taken into account 
in order to model the vibrational progression observed in experimental Auger 
electron spectra.

\section*{Supplementary Material}
See supplementary material for molecular structures, details about the basis sets used in our calculations as well as further results.

\begin{acknowledgments}
The authors thank Professors Lorenz S. Cederbaum and Anna I. Krylov as well 
as Drs. Axel Molle and Wojciech Skomorowski for helpful discussions. 
We also thank the anonymous reviewers for their comments 
that improved the manuscript significantly. \mbox{T.-C.}~J. gratefully 
acknowledges funding from the European Research Council (ERC) under the 
European Union's Horizon 2020 research and innovation program (Grant 
Agreement No. 851766). F. M. is grateful for a Kekul\'e fellowship 
(K 208/24) by the Fonds der Chemischen Industrie.
\end{acknowledgments}

\section*{Data Availability Statement}

The data that support the findings of this study are available within the article and its supplementary material.

\section*{Author Declarations}
The authors have no conflicts to disclose. 

\bibliography{main}% Produces the bibliography via BibTeX.

\end{document}